\documentclass[journal,10pt,doublecolumn]{IEEEtran}
\usepackage{mathpple}
\usepackage{times}

\usepackage{amsmath}  % Define \boldsymbol (in amsbsy too) and align
\usepackage{amssymb}  % Define \mathbb (in amsfonts too)
\usepackage{mathrsfs} % Define \mathscr, a script font
\usepackage{theorem}  % Helps in rendering theorems, etc.
\usepackage{cite}     % Gives multiple references as intervals
\usepackage{comment}  % Comments out with \begin{comment} ... \end{comment}
\usepackage[colorlinks,
linkcolor=black,
anchorcolor=black,
citecolor=black
]{hyperref}
\usepackage{upref}
\usepackage{amsfonts}
\usepackage{dsfont}
\usepackage{verbatim}
\usepackage{algorithm}
\usepackage{algorithmic}
\usepackage[dvipsnames,usenames]{color}
\usepackage{enumerate}

\usepackage{graphicx}
\usepackage{subfigure}

\usepackage{latexsym}

\usepackage{color}
\usepackage{multirow}
\usepackage{footnote}
\usepackage{booktabs}
\usepackage{cite}
\usepackage{hhline}
\usepackage{gensymb}
\usepackage{tikz}
\usepackage{graphicx}
\usepackage{xcolor}
\usepackage{listings}
\usepackage[
top    = 0.75in,
bottom = 1.05in,
left   = 0.63in,
right  = 0.63in]{geometry}
\usepackage{caption}
\usepackage{romannum}
\newcommand{\RNum}[1]{\textup{\uppercase\expandafter{\romannumeral#1\relax}} 
	
}
\newcommand{\be}[1]{\begin{equation}\label{#1}}
\newcommand{\ee}{\end{equation}}

\newcommand{\bc}{\begin{center}}
	\newcommand{\ec}{\end{center}}

\newcommand{\ceilenv}[1]{\left\lceil #1 \right\rceil}

\newcommand{\qed}{\hfill$\Box$}
\renewcommand{\leq}{\leqslant}

\renewcommand{\geq}{\geqslant}

\newcommand{\Cref}[1]{Co\-rol\-la\-ry\,\ref{#1}}

\theoremstyle{plain} \theorembodyfont{\normalfont\slshape}

\newtheorem{thm}{Theorem$\!$}
\newenvironment{theorem}{\begin{thm}\hspace*{-1ex}{\bf.}}{\end{thm}}

\newtheorem{prop}[thm]{Proposition$\!$}

\newtheorem{lem}[thm]{Lemma$\!$}
\newenvironment{lemma}{\begin{lem}\hspace*{-1ex}{\bf.}}{\end{lem}}

\newtheorem{cor}[thm]{Corollary$\!$}

\newtheorem{prob}[thm]{Problem$\!$}

\newtheorem{defi}[thm]{Definition$\!$}
\newenvironment{definition}{\begin{defi}\hspace*{-1ex}{\bf.}}{\end{defi}}

\theorembodyfont{\normalfont}

\newtheorem{exam}{Example$\!$}
\newenvironment{example}{\begin{exam}\hspace*{-1ex}{\bf .}}{\end{exam}}

\newtheorem{remrk}{Remark$\!$}
\newenvironment{remark}{\begin{remrk}\hspace*{-1ex}{\bf .}}{\end{remrk}}

\definecolor{Codecolor}{named}{White}  %{Tan}
\begin{document}
	\title{Rate-Constrained Shaping Codes for Structured Sources}
	\author{\large Yi Liu,~\IEEEmembership{Student Member,~IEEE,} Pengfei Huang,~\IEEEmembership{Student Member,~IEEE,}\\ Alexander W. Bergman, \IEEEauthorblockN{and Paul H. Siegel,~\IEEEmembership{Fellow,~IEEE}}\\
		%	\IEEEauthorblockA{\IEEEauthorrefmark{1}Electrical and Computer Engineering Dept., University of California, San Diego, La Jolla, CA 92093 U.S.A \\}
		%	{\it  \{yil333, pehuang, awbergma, psiegel\}@ucsd.edu}  
		\thanks{
			Portions of this paper were presented at the 8th Annual Non-Volatile Memories Workshop, La Jolla, CA, March 12--14, 2017, the IEEE International Symposium on Information Theory, Aachen, Germany, June 25-30, 2017, and the 9th Annual Non-Volatile Memories Workshop, La Jolla, CA, March 11--13, 2018.
			
			The authors are with the Department of Electrical and Computer Engineering, University of California, San Diego, La Jolla, CA 92093, USA, and also with
			the Center for Memory and Recording Research, University of California, San Diego, La Jolla, CA 92093-0401, USA (e-mail: \texttt{\{yil333, pehuang, awbergma, psiegel\}@ucsd.edu}).
		}
	}
	\maketitle
	\pagenumbering{arabic}
	\thispagestyle{plain}
	\pagestyle{plain}
	\begin{abstract}
		%%%%%%%%%%%%%%%%%%%%%%%%%%%%%%%%%%%%%%%%%%%%%%%%%%%%%%
		Shaping codes are used to encode information for use on channels with cost constraints. Applications include data transmission with a power constraint and, more recently, data storage on flash memories with a constraint on memory cell wear.  In the latter application, system requirements often impose a rate constraint.  In this paper, we study rate-constrained fixed-to-variable length shaping  codes for noiseless, memoryless costly channels and general i.i.d. sources.  
		The analysis relies on the theory of word-valued sources. We establish a relationship between the code expansion factor and minimum average symbol cost.  We then determine the expansion factor that minimizes the average cost per source symbol (total cost), corresponding to a conventional optimal source code with cost.  An equivalence is established between codes minimizing average symbol cost and codes minimizing total cost, and a separation theorem is proved, showing that optimal shaping can be achieved by a concatenation of optimal compression and optimal shaping for a uniform i.i.d. source. Shaping codes often incorporate, either explicitly or implicitly, some form of non-equiprobable signaling. We use our results to further explore the connections between shaping codes and codes that map a sequence of i.i.d. source symbols into an output sequence of symbols that are approximately independent and distributed according to a specified target distribution, such as  distribution matching (DM) codes. 
		Optimal DM codes are characterized in terms of a new performance measure -   generalized expansion factor (GEF)  - motivated by the costly channel perspective. The GEF is used to study  DM codes that minimize informational divergence and normalized informational divergence. 
		%%%%%%%%%%%%%%%%%%%%%%%%%%%%%%%%%%%%%%%%%%
	\end{abstract}

	%\today
	
	%%%%%%%%%%%%%%%%%%%%%%%%%%%%%%%%%%%%%%%%%%%%%%%%%%%%%%%%%%%%%%%%
	%%%%%%%%%%%%%%%%%%%%%%%%%%%%%%%%%%%%%%%%%%%%%%%%%%%%%%%%%%%%%%%%
	\section{Introduction}
	
	Shaping codes are used to encode information for use on channels with a cost constraint. A prominent application is in data transmission with a power constraint, where constellation shaping is achieved by addressing into a suitably designed multidimensional constellation or, equivalently, by incorporating, either explicitly or implicitly, some form of non-equiprobable signaling.  An excellent reference on this topic is Fischer~\cite{Fischer}.  
	
	More recently, shaping codes have been proposed for use in data storage on flash memories subject to a constraint on memory cell wear.  In that application, storage system requirements often impose a rate constraint, and the data source may be structured, rather than unconstrained. Motivated by this scenario, this paper investigates information-theoretic properties and design of rate-constrained fixed-to-variable length shaping  codes for noiseless, memoryless costly channels and general i.i.d. sources.  The analysis relies on the theory of word-valued sources developed in Nishiara and Morita~\cite{NishiaraMorita}.  Our primary interest is in the design of codes that minimize the average cost per code symbol for a given rate, or expansion factor, which we refer to as the \emph{type-\Romannum{1} shaping problem}. We also consider the well-studied problem of designing codes that minimize average cost per code symbol, or \emph{total cost}, which we refer to as the  \emph{type-\Romannum{2} shaping problem}.  
	
	The word-valued source analysis provides a natural link between shaping codes and codes that efficiently map a sequence of i.i.d. source symbols into an output sequence of symbols that are approximately independent and distributed according to a specified target distribution. Such codes have been studied in the context of random number generating source codes by Han and Uchida~\cite{HanUchida} and as distribution matching (DM) codes by B\"{o}cherer and Mathar~\cite{BochererMatharDM}, B\"{o}cherer~\cite{BochererThesis}, Amjad and B\"{o}cherer~\cite{AmjadBocherer},  B\"{o}cherer and Amjad~\cite{BochererAmjad}, Schulte and B\"{o}cherer~\cite{SchulteBocherer} and Schulte and Steiner~\cite{SchulteSteiner}.  Our shaping code analysis suggests a new performance measure - generalized expansion factor (GEF) - for fixed-to-variable length DM codes which we use to study codes that minimize informational divergence and normalized informational divergence from a shaping code perspective. 
	
	There is a substantial literature on shaping codes and, more recently, a body of work relating to DM codes.  Therefore, before summarizing our results in more detail, we provide a brief review of relevant work in both of these areas as a framework for our contributions. 
	
	%%%%%%%%%%%%%%%%%%%%%%%%%%%%%%%%%%%%%%%%%
	\subsection{Shaping Codes}
	\subsubsection{Codes minimizing total cost}
	The problem of coding for noiseless costly channels, or source coding with unequal symbol costs, traces its conceptual origins to Shannon's 1948 paper that launched the study of information theory~\cite{Shannon}. In that paper, Shannon considered the problem of transmitting information over a telegraph channel. The channel symbols -- dots and dashes -- have different time durations, which can be interpreted as transmission costs. Shannon determined the symbol probabilities that maximize the data transmission rate with integer symbol costs. This result was then generalized to arbitrary positive symbol costs by Krause~\cite{Krause} and Csisz\'{a}r~\cite{Csiszar1969}. 
	
	Several researchers have considered the problem of designing codes for costly channels with an i.i.d. source.  Most of this work has emphasized construction of codes that minimize average cost per source symbol, which we refer to as \emph{total cost}, without an explicit rate constraint. 
	In Karp~\cite{Karp}, costly channel coding was studied from an algebraic perspective, and the problem of designing a shaping code to minimize total cost was recast  as an integer programming problem. However, this code design approach is not computationally practical, and the algorithm proposed to reduce the complexity will result in sub-optimal results. In Golin et al.~\cite{Golin}, a dynamic programming solution for this integer programming problem was proposed, providing a polynomial time bound on complexity. 
	Other approaches using  tree-based constructions were proposed in~\cite{Krause}, Melhorn~\cite{Mehlhorn}, and  Csisz\'{a}r and K\"{o}rner~\cite{CsiszarKorner}. They all constructed asymptotically optimal prefix-free variable-length shaping codes.
	A universal coding scheme based on types was also introduced in ~\cite{CsiszarKorner}. 
	
	A special case, corresponding to a uniform i.i.d. source  in which all codewords are equally likely to occur, was studied by Varn~\cite{Varn}, who proposed a variable-length code construction that minimizes the average codeword cost for a fixed codebook size. This coding technique was then incorporated into a universal coding scheme in Iwata~\cite{Iwata}, which combines LZ78 compression with Varn coding. Later in this paper, we generalize the Iwata scheme, which can be viewed as an embodiment of a separation theorem proved in Section~\ref{OptimalCodes}, and further explore properties and applications of Varn codes.
	
	A generalization of Huffman coding for unequal symbol costs was proposed in Gilbert ~\cite{Gilbert}. In Guazzo~\cite{Guazzo}, practical arithmetic coding was introduced. This coding technique was then generalized by Savari and Gallager~\cite{SavariGalleger} and its properties, such as optimality and coding delay, were analyzed. However, the analysis is based on infinite precision arithmetic coding, which cannot be realized in practice.  
	
	In B\"{o}cherer and Mathar~\cite{BochererMatharDM} and  B\"{o}cherer~\cite{BochererThesis}, a  
	variable-to-fixed length code construction called geometric Huffman coding  was used to design codes for an i.i.d. uniform source that asymptotically minimize the total cost of a noiseless channel with unequal symbol durations. (This construction matches codeword probabilities to dyadic symbol distributions that optimally approximate the optimal symbol distribution.)
	
	We emphasize that all of the codes mentioned above considered the problem of minimizing cost per source symbol, i.e., total cost, with no explicit consideration of rate. The dependence of total cost on code rate was not thoroughly investigated.
	
	%%%%%%%%%%%%%%%%%%%%%%%%%%%%%%%%%%%%%%%%%%
	\subsubsection{Rate-constrained codes minimizing average cost}
	The problem of designing rate-constrained codes for costly channels has received less attention. 
	The maximum entropy of a stationary Markov chain on a finite-state channel with associated symbol/transition costs, along with the entropy-maximizing symbol/transition probabilities, can be found in McEliece and Rodemich~\cite{McElieceRodemich}, Justesen and H{\o}holdt~\cite{Justesen}, and Khandekar, McEliece, and Rodemich~\cite{Khandekar}. In McEliece~\cite{McElieceBook} and, later,  B\"{o}cherer~\cite{BochererThesis}, the special case corresponding to a memoryless channel is addressed. 
	
	B\"{o}cherer and Mathar~\cite{BochererMatharDM} and  B\"{o}cherer~\cite{BochererThesis} apply the geometric Huffman coding approach to design variable-to-fixed length codes that match codeword probabilities to dyadic symbol distributions that approximate the entropy-maximizing probability mass function for    memoryless costly channels subject to an average cost constraint, thereby asymptotically achieving the maximum rate. 
	
	The state-splitting algorithm~\cite{ACH}, which was developed to construct finite-state codes for constrained channels, has been extended for application to construction of codes for costly channels. Heegard, Marcus, and Siegel~\cite{HMS} studied a class of channels with average runlength constraints, which represent a special case of noiseless channels with a cost constraint.
	They constructed variable-to-variable length synchronous codes using state-splitting techniques adapted for channels with variable-length symbols.   
	Khayrallah and Neuhoff~\cite{KhayNeu} and McLaughlin and Khayrallah~\cite{McLKhay}  construct  fixed-to-fixed length and variable-to-fixed length codes based on state-splitting methods for  magnetic recording and constellation shaping applications.  Krachkovsky et al.~\cite{Krach} determine a costly channel model matched to a Markov source and construct corresponding codes using enumerative techniques for application to transmission over an intersymbol-interference channel. All of these works strive to construct codes that come close to the capacity-cost functions originally presented in~\cite{McElieceBook},\cite{McElieceRodemich}, and~\cite{Justesen}.
	
	Other recent work relating to this problem has been motivated by non-volatile memory applications, so we briefly describe the corresponding costly channel model. 
	NAND flash memory  uses floating-gate transistors, commonly referred to as \emph{cells}, to store information in the form of different cell voltage levels.   The flash memory cells gradually wear out with repeated writing and erasing, referred to as program/erase cycling, and the damage caused by the cycling is dependent on the programmed voltage levels~\cite{LiuNVMW2016},~\cite{LiuSieGC16}. The costly channel model associates to each cell voltage level a wear cost reflecting the extent of the damage induced by writing that level. 
	
	Recently, in~\cite{Jagmohan}, Jagmohan et al. proposed \textit{endurance coding}, intended for  shaping of programmed data for flash memories. For a given cost model and a specified target code rate, the optimal distribution of cell levels that minimizes the average cost was determined analytically, reproducing the results in the references cited above. For single bit per cell (SLC) flash memory, with associated level costs of 0 and 1,  greedy enumerative codes that minimize the number of cells with cost 1 were designed and evaluated in terms of the rate-cost trade-off.  However, endurance coding is intended for uniform i.i.d. source data. For structured source data, which would include a general i.i.d. source,  the idea of combining source compression with endurance coding was proposed, but the relationship between the code performance and the code rate for arbitrary sources was not thoroughly studied.
	
	In Sharon et al.~\cite{Sharon}, low-complexity, rate-1, fixed-length \textit{direct shaping codes} for structured data were proposed for use on SLC flash memory. The  code construction used a greedy approach based upon an adaptively-built encoding dictionary that does not require  knowledge of the source statistics. This construction was extended to a direct shaping code compatible with two-bit per cell (MLC) flash memory operation by Liu et al. in~\cite{LiuNVMW2016}, \cite{LiuSieGC16}. However, it was proved in Liu and Siegel~\cite{LiuISIT2017} that direct shaping codes are in general suboptimal.  (Our experimenal results in Section~\ref{sec:experiment} contain  a comparison of a shaping scheme motivated by our analysis to a direct shaping code on MLC flash memory.)  
	
	\subsubsection{Summary of contributions on shaping codes}
	In this paper, our goal is to systematically study the fundamental performance limits of fixed-to-variable length shaping codes from a rate and distribution perspective.  We first use known properties of  word-valued sources to determine the symbol occurrence probability of shaping code output sequences 
	(Lemma~4).  We then derive an upper bound on the code sequence entropy rate (Lemma~5). 
	Using these results, we are able to reduce the problem of minimizing average code symbol cost subject to a constraint on the code rate to an optimization problem for an i.i.d. process. This problem can be viewed as the dual problem to the entropy-maximization problem considered in the prior literature. 
	We refer to this minimization problem as the \textit{type-\Romannum{1} shaping problem}, and we call shaping codes that achieve the minimum average cost for a given rate \textit{optimal type-\Romannum{1} shaping codes.}
	We develop a theoretical bound on the trade-off between the rate -- or more precisely, the  corresponding \emph{expansion factor} -- and the average cost  of a type-\Romannum{1} shaping code (Theorem~6).
	We then study shaping codes that minimize total cost (minimum average cost per source symbol). We refer to the problem of minimizing the total cost as the \textit{type-\Romannum{2} shaping problem} and shaping codes that achieve the minimum total cost are referred to as \textit{optimal type-\Romannum{2} shaping codes}.
	We derive the relationship between the code expansion factor and the total cost and determine the optimal expansion factor (Theorem~7). We then prove an equivalence theorem showing that an optimal type-\Romannum{1} shaping code can be realized using an optimal type-\Romannum{2} shaping code for another suitably chosen costly channel model (Theorem~8). 
	We can therefore solve the type-\Romannum{1} shaping problem using known coding techniques such as generalized Shannon-Fano codes~\cite{CsiszarKorner} . 
	A consequence of the analysis is a separation theorem for type-\Romannum{1}I shaping codes, which states that optimal shaping can be achieved by a concatenation of lossless compression and optimal shaping for a uniform i.i.d. source.  This provides an alternative architecture for implementing asymptotically optimal shaping codes using, for example, Varn codes. 
	Finally, we prove a separation theorem for type-\Romannum{1} shaping codes with given expansion factor, using a careful analysis of the behavior of the minimum average cost as a function of the expansion factor. 
	
	\subsection{Distribution Matching (DM) Codes}
	\subsubsection{Applications of DM codes to shaping}
	The application of non-equiprobable signaling in the context of coding with a cost constraint reflects the interesting interplay between shaping codes and DM-type codes (in the broad sense of codes that map an i.i.d. sequence of source symbols to an output sequence of symbols that are approximately independent and distributed according to $\{P_i\}$). 
	Beginning with the work on constellation shaping, there have been a number of applications of DM-type codes to coding for a costly channel. 
	
	In \cite{Forney},  signal constellations with non-uniform symbol probabilities were used for efficient modulation on band-limited channels. Noting the dual nature of non-equiprobable signaling and source coding,  several authors proposed the use of ``reverse'' source codes derived from, for example,  Huffman codes, Tunstall codes, and arithmetic codes  as DM codes for shaping applications. See, for example, works by Kschishang and Pasupathy~\cite{KP},  Ungerboeck~\cite{Ungerboeck}, Abrahams~\cite{Abrahams},  Baur and B\"{o}cherer\cite{Baur}.
	
	Gallager~\cite[p. 208]{Gallager} proposed a method of generating symbols with a biased distribution to be combined with linear coding as an approach to achieving capacity of an asymmetric channel. This idea was incorporated into a general scheme that can use capacity-achieving codes for symmetric channels, such as polar codes, to achieve the capacity of arbitrary discrete  memoryless asymmetric channels in Mondelli et al.~\cite{Mondelli}. 
	
	In~\cite{BochererSteinSchulte}, B\"{o}cherer et al. propose a scheme that combines DM codes (such as constant composition codes) with systematic error correction codes. This scheme can be regarded as a simplification of the bootstrap scheme in B\"{o}cherer and Mathar~\cite{BochererMatharLDPC}, which concatenates the check bits generated  by the systematic ECC encoder with the following information bits and applies a DM encoder to them. In~\cite{Mondelli}, the authors also proved that the bootstrap scheme, which they refer to as a chaining construction, can be used to achieve the capacity of any discrete memoryless asymmetric channel. 
	
	\subsubsection{DM codes with optimality measures}
	In Han~\cite{HanBook} and Visweswariah et al.~\cite{Vis1998},  it was shown that an optimal variable-length source code can be regarded as an optimal variable-length DM code for a uniform distribution. The criterion for optimality was the vanishing of a form of normalized conditional Kullback-Leibler (KL) divergence between a subset of codewords of fixed length and words generated i.i.d. with the target distribution, asymptotically in the block length. This result was further developed in Han and Uchida~\cite{HanUchida}, where an optimal variable-length source code with cost, meaning a code that minimizes total cost, was shown to be an optimal DM code. The maximum achievable rate of non-prefix-free DM codes was discussed in Uchida~\cite{Uchida}.
	
	In~\cite{BochererMatharDM}, dyadic probability mass functions with some optimality properties were used to match the capacity-achieving probability distribution of a discrete memoryless channel, and variable-to-fixed length geometric Huffman codes, mentioned earlier, were used as DM codes.   Normalized informational divergence --  defined as the  KL-divergence between a codeword probability distribution and the distribution of the codewords when generated i.i.d. by the target distribution, normalized by the codeword length -- was introduced as the DM code optimality measure.    It was then proved that geometric Huffman coding  is asymptotically optimal, in the sense that the normalized informational divergence converges to zero as the codeword length increases. 
	Other fixed-length DM codes with vanishing normalized informational divergence were presented in Ramabadran~\cite{Ramabadran} and~\cite{SchulteBocherer}. 
	
	Constellation shaping techniques have also been adapted for use in DM coding. For example, a DM code using shell mapping was presented in~\cite{SchulteSteiner} and a DM code using trellis shaping was presented by Gultekin et al.~\cite{Gultekin}. 
	
	In~\cite{AmjadBocherer}, the notions of informational divergence and normalized information divergence were extended to measure the performance of fixed-to-variable length codes. Optimality of complete Tunstall code trees with respect to minimizing informational divergence was proved, a result we extend in Section~\ref{sec::compare}. An efficient algorithm for finding binary DM codes that minimize the normalized informational divergence, based on an iterative adaptation of binary Tunstall coding, was presented, and asymptotic optimality with increasing block length was established. In~\cite{BochererAmjad} the relationship between normalized information divergence of a DM code and its rate was studied, a topic that we further address in Section~\ref{sec::compare}. 
	
	\subsubsection{Summary of contributions on DM codes}
	In this paper, we systematically study the problem of designing optimal fixed-to-variable length, prefix-free DM codes from the perspective of word-valued sources and shaping codes.  The degree of distribution matching is measured by  the KL-divergence between the distribution on word-valued source output sequences and the distribution on those sequences generated i.i.d. according to the target distribution. Vanishing asymptotic normalized KL-divergence at the sequence level, suggested by the approach in~\cite{NishiaraMorita} and also studied by Soriaga~\cite{Soriaga}, is used as the criterion for optimality.  We first characterize the expansion factor of an optimal DM code for a general i.i.d. source (Theorem 12).  We then show that an optimal type-\Romannum{2} shaping code for a cost model determined by the negative logarithm of a target distribution   is an optimal DM code for that distribution (Theorem 13). (This ``self-information'' cost model was also used in~\cite{SchulteSteiner} to design information divergence optimal fixed-to-fixed DM codes using shell mapping.) 
	
	The connection between   shaping codes and DM codes suggests another measure for evaluating DM code performance, which we refer to as  \textit{generalized expansion factor} (GEF). We establish a lower bound on the generalized expansion factor, and show that a code that achieves the lower bound is an optimal DM code (Theorem 15). This implies that Varn codes are asymptotically optimal DM codes for a uniform i.i.d. source. Using the GEF, we also extend the separation theorem of shaping codes to DM codes (Theorem 16). 
	
	Finally, we discuss relationships between different DM code performance measures. We show that for a DM code with fixed codebook size, minimizing the GEF  is equivalent to minimizing the informational divergence introduced in~\cite{AmjadBocherer}, leading to the conclusion that Varn codes designed for the appropriate cost model minimize informational divergence (Theorem 17), generalizing a result for binary Tunstall codes in~\cite{AmjadBocherer}.  We also give an explicit description of the relationship between the normalized informational divergence of a DM code and its expansion factor (Theorem 18), refining a bound in~\cite{BochererAmjad}.  
	
	\subsection{Organization of the Paper}
	The remainder of the paper is organized as follows.
	In Section \ref{Preliminaries}, we use known properties of  word-valued sources to determine the symbol occurrence probability of shaping code output sequences and the lower bound  on the symbol distribution entropy. In Section~\ref{OptimalCodes}, we analyze the distribution, cost, and rate properties of fixed-to-variable length shaping codes. The analysis is then used to prove the equivalence theorem and separation theorem. In Section~\ref{sec:DM}, we establish the equivalence between optimal distribution matching codes and optimal shaping codes. Section~\ref{sec:gef} introduces the generalized expansion factor and proves the separation theorem for DM codes. Section~\ref{sec::compare} compares different DM code performance measures. In Section~\ref{sec:experiment}, we apply a shaping scheme motivated by our theoretical results to a multilevel flash memory. and we show simulation results illustrating the application of Varn codes to DM coding. Section~\ref{sec:conclude} concludes the paper.
	
	%%%%%%%%%%%%%%%%%%%%%%%%%%%%%%%%%%%%%%%%%%%%%%%%%%%%%%%%%%%%%
	%%%%%%%%%%%%%%%%%%%%%%%%%%%%%%%%%%%%%%%%%%%%%%%%%%%%%%%%%%%%%
	\section{Information-theoretic Preliminaries}
	\label{Preliminaries}
	%%%%%%%%%%%%%%%%%%%%%%%%%%%%%%%%%%%%%%%%%%%%%%%%%%%%%%%%%%%%%
	%%%%%%%%%%%%%%%%%%%%%%%%%%%%%%%%%%%%%%%%%%%%%%%%%%%%%%%%%%%%%
	\subsection{Basic Model}
	First, we fix some notation.  Let $\mathbf{X} = X_1X_2\ldots$, where $X_i  \sim X$ for all $i$,  be an i.i.d. source with alphabet $\mathcal{X}=\{\alpha_1,\ldots,\alpha_u\}$. We use $|\mathcal{X}|$ to denote the size of the alphabet and use $P(x^*)$ to denote the probability of any finite sequence $x^*$. Let $\mathcal{Y}=\{\beta_1,\ldots , \beta_v\}$ be an alphabet and $\mathcal{Y}^*$ be the set of all finite sequences over $\mathcal{Y}$, including the null string $\lambda$ of length 0. Each $\beta_i$ is associated with a   cost $C_i$. Without loss of generality, we assume that $0\leq  C_1\leq C_2\leq \ldots C_v$, and we also assume that there exists at least one pair of costs, $C_i$ and $C_j$, such that $C_i\neq C_j$. We use a cost vector $\mathcal{C}=[C_1, C_2, \ldots, C_v]$ to represent the   cost associated with alphabet $\mathcal{Y}$.
	
	A general shaping code is defined as a prefix-free variable-length mapping $\phi: \mathcal{X}^q \rightarrow \mathcal{Y}^*$ which maps a length-$q$ data word $x_1^q$ to a variable-length codeword $y^*$. We use  
	$\mathbf{Y}$ to denote the process $\phi(\mathbf{X}^q)$,
	where $\mathbf{X}^q $ is the vector process  $X_1^q, X_{q+1}^{2q}, \ldots$. The entropy rate of the process $\mathbf{Y}$ is
	
	\begin{equation}
	H(\mathbf{Y})=\lim_{n\rightarrow \infty}\frac{1}{n}H(Y_1Y_2\ldots Y_n).
	\end{equation}
	
	We denote the length of a codeword $\phi(x_1^q)$ by $L(\phi(x_1^q))$ and the expected length of  codewords generated by a sequence of length-$q$ source words is given by 
	%\vspace{-1ex}
	\begin{equation}
	E(L)=\sum_{x_1^q\in \mathcal{X}^q} P(x_1^q)L(\phi(x_1^q)).
	\end{equation}
	The \textit{expansion factor} is defined as the ratio of the  expected codeword length to the length of the input source word, namely 
	\begin{equation}
	f=E[L]/q.
	\end{equation}
	\begin{remark}
		Endurance codes and direct shaping codes can be treated as special cases of this general class of shaping codes. 
		Endurance codes are used when the source    has a uniform i.i.d. distribution, with entropy rate  $H(\mathbf{X})= \log_2 |\mathcal{X}|$. A length-$m$ direct shaping code is a shaping code with $q=1$, $f=1$, where both 
		$\mathbf{X}$ and $\mathbf{Y}$ have   alphabet size $2^m$.
		\qed
	\end{remark}
	
	The pair $\mathbf{X}$ and $\phi$ form a word valued source, as defined in\cite{NishiaraMorita}. The following theorem, proved in\cite{NishiaraMorita},  gives the entropy rate of the shaping code $\phi(\mathbf{X}^q)$.
	\begin{theorem}
		\label{Hiroyoshitheorem}
		For a prefix-free variable-length code $\mathbf{Y} = \phi(\mathbf{X}^q)$ such that $H(\mathbf{X}^q)<\infty$ and $E(L)<\infty$, the entropy rate of the encoder output satisfies
		\begin{equation}
		H(\mathbf{Y})=\frac{H(\mathbf{X}^q)}{E(L)}=\frac{qH(\mathbf{X})}{E(L)}.   
		\end{equation}\qed
	\end{theorem}
	
	%%%%%%%%%%%%%%%%%%%%%%%%%%%%%%%%%%%%%%%%%%%%%%%%%%%%%%%%%%%%%
	%%%%%%%%%%%%%%%%%%%%%%%%%%%%%%%%%%%%%%%%%%%%%%%%%%%%%%%%%%%%%
	%\vspace{-2em}
	\subsection{Asymptotic Symbol Occurrence Probability}
	
	For simplicity and without loss of generality, we assume $q=1$. The mapping is $\phi: \mathcal{X}\rightarrow \mathcal{Y}^*$.
	Let $y_1^l$ denote the first $l$ symbols of $\phi(\mathbf{X})$. We assume the   cost is independent and additive, so the   cost of sequence $y_1^l$ can be expressed as 
	\begin{equation}
	W(y_i^l) = \sum_{i=1}^v N_i(y_1^l)C_i
	\vspace{-1ex}
	\end{equation}
	where $N_i(y_1^l)$ stands for the number of occurrences of $\beta_i$ in sequence $y_1^l$.  The   cost per code symbol is  therefore $\sum_i N_i(y_1^l)C_i/l$.
	Let 
	\begin{equation}
	Q(y_1^l)=Pr\{Y_1^l =y_1^l\}
	%\vspace{-1ex}
	\end{equation}
	denote the probability distribution of $Y_1^l$.
	The expected   cost per symbol of a length-$l$ shaping code sequence is 
	\begin{equation}
	\vspace{-1ex}
	\label{equ:averagewearcost1}
	\begin{split}
	W_l &= \sum_{y_1^l \in \mathcal{Y}^l}   Q(y_1^l)W(y_i^l) /l\\
	& = \sum_{i=1}^v \sum_{y_1^l \in \mathcal{Y}^l} Q(y_1^l) N_i(y_1^l) C_i /l.
	\end{split}
	\vspace{-1ex}
	\end{equation}
	The asymptotic expected   cost per symbol, or \textit{average   cost} of a shaping code is
	\vspace{-1ex}
	\begin{equation}
	\label{equ:averagewearcost2}
	A(\phi(\mathbf{X}))=\lim_{l \rightarrow \infty}W_l .
	%= \lim_{l \rightarrow \infty} \sum_{i=1}^v \sum_{y_1^l \in \mathcal{Y}^l} Q(y_1^l)N_i(y_1^l) C_i /l .
	\end{equation}
	Let 
	\begin{equation}
	\label{equ:averagewearcost3}
	\hat{P_i}  = \lim_{l\rightarrow \infty} \sum_{y_1^l} Q(y_1^l)N_i(y_1^l)/l\\  =\lim_{l\rightarrow \infty} \frac{E(N_i(Y_1^l))}{l}
	\end{equation}
	be the asymptotic probability of occurrence of $\beta_i$. Then the average cost of a shaping code can be expressed as 
	\begin{equation}
	A(\phi(\mathbf{X})) = \sum_i \hat{P_i} C_i.
	\end{equation}
	
	In the rest of this subsection, we will show how to calculate $\hat{P_i}$. Define the prefix operator $\pi$ as $y_1^n\pi^i = y_1^{n-i}$ for $0\leq i < n$ and $y_1^n\pi^i = \lambda$ for $i\geq n$. Let $\pi\{y^*\}$ denote the set of all the prefixes of a sequence $y^*$.
	We denote by $\mathcal{G}_\phi(y_1^l)$ the set of all sequences $x^*\in \mathcal{X}^*$ such that $y_1^l$ is a prefix of $\phi(x^*)$ but not of $\phi(x^*\pi)$. That is, 
	\begin{equation}
	\mathcal{G}_{\phi}(y_1^l) = \{x^*\in \mathcal{X}^*|y_1^l\in\pi\{\phi(x^*)\}\wedge|\phi(x^*\pi)|<l\}
	\end{equation}
	and the distribution of $y_1^l$  can be expressed as
	\begin{equation}
	Q(y_1^l) = \sum_{x^*\in \mathcal{G}_{\phi}(y_1^l)} P(x^*).
	\end{equation}
	We define by $M_l$ the minimum length of a sequence $x_1^n$ such that $|\phi(x_1^n)|\geq l$ and let $S_{M_l}$ be the length of $\phi(x_1^{M_l})$.  Note that 
	\begin{equation}
	S_{M_l-1}<l\leq S_{M_l}.
	\end{equation}
	According to \cite{NishiaraMorita}, the random variable $M_l$ satisfies the property of being a \textit{stopping rule} for the sequence of i.i.d. random variables $\{\phi(X^\infty)\}$. Wald's equality \cite{Wald} then implies that 
	\begin{equation}
	\label{equ:walds}
	E(N_i(\phi(X_1^{M_l})))=E(N_i(\phi(X)))E(M_l).
	\vspace{-1ex}
	\end{equation}
	The following two lemmas were proved in \cite{NishiaraMorita}.
	\vspace{-1ex}
	\begin{lemma}
		\label{lemma:11}
		Given a nonnegative-valued function $f$, let $F_i=f(X_i)$. If $E(F)<\infty$, then
		\begin{equation}
		\lim_{l\rightarrow \infty}\frac{E(F_{M_l})}{l}=0.
		\end{equation}\qed
	\end{lemma}
	
	\begin{remark}
		The previous lemma is not obvious, because even when  $E(F)<\infty$, 
		$E(F_{M_l})$ is not necessarily equal to $E(F)$. \qed
	\end{remark}
	
	\begin{lemma}
		\label{lemma::wald} 
		If $E[L]<\infty$, then
		\begin{equation}
		\lim_{l\rightarrow \infty}\frac{E[M_l]}{l} = \frac{1}{E(L)}.
		\end{equation}\qed
	\end{lemma}

	Using these results, we derive a lemma which tells us how to calculate the asymptotic occurrence probability of the encoder output process $\mathbf{Y}$.
	
	\vspace{-1em}
	\begin{lemma}
		\label{lemma:10}
		For a prefix-free variable-length code $\phi: \mathcal{X}^q \rightarrow \mathcal{Y}^*$ such that $E(N_i(\phi(X^q))) < \infty$ for all symbols $\beta_i$ and ${E(L)<\infty}$, the asymptotic probability of occurrence $\hat{P_i}$  of $\beta_i$ is given by 
		\begin{equation}
		\hat{P_i}=E(N_i(\phi(X^q)))\frac{1}{E(L)}.
		\end{equation} %~\qed
	\end{lemma}
	\begin{IEEEproof}
		See Appendix \ref{appen:1}.
	\end{IEEEproof}
	It is easy to check that $\sum_i \hat{P_i} = 1$, so this distribution is well defined. 
	
	%%%%%%%%%%%%%%%%%%%%%%%%%%%%%%%%%%%%%%%%%%%%%%%%%
	%%%%%%%%%%%%%%%%%%%%%%%%%%%%%%%%%%%%%%%%%%%%%%%%%
	\subsection{Lower Bound on Symbol Distribution Entropy}
	Consider a prefix-free variable-length code $\phi$ as in Lemma~\ref{lemma:10}. 
	Let $\hat{Y}_1^l$ denote an i.i.d. sequence of length $l$ and with distribution $\{\hat{P_i}\}$. The probability of a length-$l$ sequence $y_1^l$ with respect to this distribution is  $\hat{P}(y_1^l) = \prod_i \hat{P_i}^{N_i(y_1^l)}$. The Kullback-Leibler (KL) divergence (also known as the  KL-distance or relative entropy)~\cite{Cover} is a measure of the inefficiency caused by an approximation. The  KL-divergence between $Y_1^l$ and $\hat{Y}_1^l$ is
	\begin{equation}
	D(Y_1^l || \hat{Y}_1^l) = \sum_{y_1^l\in \mathcal{Y}^l} Q(y_1^l) \log_2 \frac{Q(y_1^l)}{\hat{P}(y_1^l)}. 
	\end{equation}
	
	%Using the fact that $D(Y_1^l || \hat{Y}_1^l) \geq 0$, we have the following lemma.
	The following lemma provides a lower bound on the symbol distribution entropy.
	\begin{lemma}
		\label{upperbound_marginaldistribution}
		The entropy $H(\hat{Y}) = -\sum_i \hat{P_i}\log_2 \hat{P_i}$ is lower bounded by the entropy rate of the shaping code sequence, i.e.,
		\begin{equation}
		H(\hat{Y}) \geq H(\mathbf{Y}).
		\end{equation}
		Specifically,
		\begin{equation}
		\lim_{l\rightarrow \infty} \frac{1}{l}D(Y_1^l || \hat{Y}_1^l) = H(\hat{Y}) - H(\mathbf{Y}).
		\end{equation}%~\qed
	\end{lemma}
	\begin{IEEEproof}
		We rewrite the $D(Y_1^l || \hat{Y}_1^l)$ as 
		\begin{equation}
		\label{I_divergence1}
		\begin{split}
		&D(Y_1^l || \hat{Y}_1^l) = \sum_{y_1^l\in \mathcal{Y}^l} Q(y_1^l) \log_2 \frac{Q(y_1^l)}{\hat{P}(y_1^l)}\\
		& = \sum_{y_1^l\in \mathcal{Y}^l} Q(y_1^l) \log_2 Q(y_1^l) - \sum_{y_1^l\in \mathcal{Y}^l} Q(y_1^l) \log_2 \hat{P}(y_1^l)\\
		& = -H(Y_1^l) - \sum_{y_1^l\in \mathcal{Y}^l} Q(y_1^l) \log_2 \hat{P}(y_1^l).
		\end{split}
		\vspace{-1ex}
		\end{equation}
		The second term of the right-hand side of this equation is
		\vspace{-1ex}
		\begin{equation}
		\label{I_divergence2}
		\begin{split}
		\sum_{y_1^l\in \mathcal{Y}^l}& Q(y_1^l) \log_2 \hat{P}(y_1^l) =  \sum_{y_1^l\in \mathcal{Y}^l} Q(y_1^l) \log_2  \prod_i \hat{P_i}^{N_i(y_1^l)}\\
		& = \sum_{y_1^l\in \mathcal{Y}^l} Q(y_1^l) \sum_i {N_i(y_1^l)}\log_2   \hat{P_i}\\
		& = \sum_i \log_2   \hat{P_i}\sum_{y_1^l\in \mathcal{Y}^l} Q(y_1^l) {N_i(y_1^l)}.
		\vspace{-1ex}
		\end{split}
		\end{equation}
		Combining equations (\ref{I_divergence1}) and (\ref{I_divergence2}), we have
		\begin{equation}
		\label{I_divergence3}
		\begin{split}
		&\lim_{l\rightarrow \infty} \frac{1}{l} D(Y_1^l || \hat{Y}_1^l) \\ &=-\lim_{l\rightarrow \infty} \frac{1}{l}H(Y_1^l)  - \sum_i \log_2   \hat{P_i}\lim_{l\rightarrow \infty}\sum_{y_1^l\in \mathcal{Y}^l} \frac{Q(y_1^l) {N_i(y_1^l)}}{l}\\
		&= -H(\mathbf{Y}) - \sum_i \hat{P_i} \log_2 \hat{P_i} = H(\hat{Y}) - H(\mathbf{Y}).
		\vspace{-1ex}
		\end{split}
		\end{equation}
		Using the fact that $D(Y_1^l || \hat{Y}_1^l) \geq 0$, we have
		\begin{equation}
		H(\hat{Y})\geq H(\mathbf{Y}).
		\end{equation}
		This completes the proof.
	\end{IEEEproof}
	
	\begin{remark}
		\label{remark::iid}
		From the proof, we see that  $H(\hat{Y}) = H(\mathbf{Y})$ implies $\lim_{l\rightarrow \infty}\frac{1}{l}D(Y^l|| \hat{Y}^l)=0$. Therefore, the codeword sequence $Y_1Y_2\cdots$ approximates an i.i.d. sequence generated by $\hat{Y}$.\qed
	\end{remark}
	
	\begin{example}
		\label{example1}
		Consider a uniform i.i.d. binary source  $\mathbf{X}$ and a prefix-free variable-length code defined by the mapping  $\{00\rightarrow 000, 01 \rightarrow 001, 10 \rightarrow 01, 11\rightarrow 1 \}$. 
		The occurrence probabilities of symbols 0 and 1 are $2/3$ and $1/3$, respectively. The symbol distribution entropy is 
		\begin{equation}
		H(\hat{Y})=-\frac{1}{3}\log_2\frac{1}{3}-\frac{2}{3}\log_2\frac{2}{3}\simeq 0.9183.
		\end{equation} 
		The entropy rate of the shaping code sequence is
		\begin{equation}
		H(\mathbf{Y})= \frac{H(\mathbf{X}^2)}{E(L)}=\frac{2}{2.25} = 0.8889.
		\end{equation}
		We see that, in this case, $H(\mathbf{Y})< H(\hat{Y})$.\qed
	\end{example}
	
	%%%%%%%%%%%%%%%%%%%%%%%%%%%%%%%%%%%%%%%%%%%%%%%%%
	%%%%%%%%%%%%%%%%%%%%%%%%%%%%%%%%%%%%%%%%%%%%%%%%%
	
	\section{Optimal Shaping Codes}
	\label{OptimalCodes}
	
	%%%%%%%%%%%%%%%%%%%%%%%%%%%%%%%%%%%%%%%%%%%%%%%%%
	%%%%%%%%%%%%%%%%%%%%%%%%%%%%%%%%%%%%%%%%%%%%%%%%%
	\subsection{Cost Minimizing Probability Distribution}
	
	In this subsection, we discuss the properties of optimal shaping codes. We consider two scenarios. First, we analyze shaping codes that minimize the average cost with a given expansion factor. We then analyze shaping codes that minimize the  expected cost per source symbol, or total cost. 
	
	We refer to the first minimization problem as the \textit{type-\Romannum{1} shaping problem}, and we call shaping codes that achieve the minimum average  cost for a given expansion factor \textit{optimal type-\Romannum{1} shaping codes}. 
	The following theorem gives a lower bound on the average cost and the corresponding asymptotic symbol occurrence probabilities.
	
	\vspace{-0.5em}
	\begin{theorem}\label{performance:opt}
		Given the source $\mathbf{X}$ and cost vector $\mathcal{C}$, the average   cost of a type-\Romannum{1} shaping code $\phi: \mathcal{X}^q \rightarrow \mathcal{Y}^*$ with expansion factor $f$ is lower bounded by $\sum_{i}\hat{P}_iC_i$, with 
		\begin{equation}
		\hat{P}_i=\frac{1}{N}2^{-\mu C_i},
		\end{equation} 
		where $N$ is a normalization constant such that $\sum_i \hat{P}_i =1$ and $\mu$ is a non-negative constant such that $\sum_i -\hat{P}_i \log \hat{P}_i=H(\mathbf{X})/f$.%\qed
	\end{theorem}
	
	%\vspace{-2em}
	\begin{IEEEproof}
		From Theorem~\ref{Hiroyoshitheorem} and Lemma~\ref{upperbound_marginaldistribution}, we see that, for a shaping code $\phi$ with expansion factor $f$, the following inequality holds:
		\begin{equation}
		H(\hat{Y})\geq H(\mathbf{Y})= \frac{qH(\mathbf{X})}{E(L)} = \frac{H(\mathbf{X})}{f}.
		\end{equation}
		To calculate the minimum possible average cost, we must solve the optimization problem:
		\begin{equation}\label{equ:optimize}
		\begin{aligned}
		& \underset{\hat{P}_i}{\text{minimize}}
		& & \sum_{i}\hat{P}_iC_i\\
		& \text{subject to}
		& & H(\hat{Y})\geq \frac{H(\mathbf{X})}{f}\\
		&  && \sum_{i}\hat{P}_i=1.
		\end{aligned}
		\vspace{-0.8em}
		\end{equation}
		
		We divide this optimization problem into two parts. First, we fix $H(\hat{Y})$ and find the optimal symbol occurrence probabilities. Then we find the optimal $H(\hat{Y})$ to minimize the average   cost. The optimization problem then becomes
		\begin{equation}\label{equ:optimizetwostep}
		\begin{aligned}
		& \underset{H(\hat{Y})}{\text{minimize}}\quad \underset{\hat{P}_i}{\text{minimize}}
		& & \sum_{i}\hat{P}_iC_i\\
		& \text{subject to}
		& & H(\hat{Y})\geq \frac{H(\mathbf{X})}{f}\\
		&  && \sum_{i}\hat{P}_i=1.
		\end{aligned}
		\vspace{-0.8em}
		\end{equation}
		If we fix $H(\hat{Y})$, we can solve the optimization problem by using the method of Lagrange multipliers. The solution is
		\vspace{-0.3em}
		\begin{equation}
		\hat{P}_i=\frac{1}{N}2^{-\mu C_i}
		\vspace{-0.5em}
		\end{equation}
		where $N=\sum_i 2^{-\mu C_i}$ is a normalization constant and $\mu$ is a non-negative constant such that 
		\vspace{-0.3em}
		\begin{equation}
		\label{equ:relation}
		H(\hat{Y})=\sum_{i}-\hat{P}_i\log_2 \hat{P}_i.\vspace{-0.5em}
		\end{equation}	
		Note that $\mu = 0$ if and only if $H(\hat{Y})=\log_2|\mathcal{Y}|$.
		For simplicity, let $h$ denote $H(\hat{Y})$.
		Then $\mu$ and $N$ are functions of $h$, which we denote by $N\stackrel{\mathrm{def}}{=}N(h)$ and $\mu\stackrel{\mathrm{def}}{=}\mu(h)$, respectively.   Let $C(h)=\sum_i \frac{C_i}{N(h)}2^{-\mu(h)C_i}$ be the minimum cost, given that $h\geq \frac{H(\mathbf{X})}{f}$. From (\ref{equ:relation}), we see that
		\vspace{-0.3em}
		\begin{equation}
		\label{equ::mcliece}
		C(h)=\frac{h-\log_2 N}{\mu}\quad \text{when $\mu>0$}.
		\vspace{-0.3em}
		\end{equation}
		%%%%%%%%%%%%%%%%%%%%%%%%%%%%%%%%%%%%%
		The optimization problem we have reduced to here, minimizing the average cost of a probability mass function  subject to a lower bound on entropy, is dual to the problem considered in prior work such as~\cite[Problem 1.8]{McElieceBook} and ~\cite[Sec. 5.2]{BochererThesis}, which is a special case of results in~\cite{McElieceRodemich}, \cite{Justesen}, and \cite{Khandekar}. 
		The relationship between entropy rate and average cost discussed in these papers has the same functional form as  (\ref{equ::mcliece}). We can apply the analysis in~\cite[Sec. 5.2]{BochererThesis}, to conclude that
		\begin{equation}
		\label{equ::averagecostandh}
		\frac{\mathrm{d}h}{\mathrm{d}C} = \mu \Rightarrow \frac{\mathrm{d}C}{\mathrm{d}h} > 0\quad \text{when $\mu > 0$}.
		\end{equation}
		Therefore, the minimum cost for a shaping code with expansion factor $f$ is achieved when  $h=H(\hat{Y})= \frac{H(X)}{f}$.
		
		Note that  we have minimized average cost by optimizing the asymptotic symbol occurrence probability $\hat{P_i}$ of a prefix-free variable-length mapping whose output entropy rate is fixed, without consideration of whether the output sequence coincides with an i.i.d. sequence.
		%%%%%%%%%%%%%%%%%%%%%%%%%%%%%%%%%%%%%%%%
	\end{IEEEproof}
	
	\begin{remark}
		\label{rmk::compareendurance}
		If the source $\mathbf{X}$ has a uniform distribution, then $\mu$ satisfies  $-f\sum_i \hat{P}_i \log \hat{P}_i=\log_2 |\mathcal{X}|$. Thus, we recover the result in \cite{Jagmohan} characterizing endurance codes with minimum average   cost.\qed
	\end{remark}
	
	\begin{remark}
		When the minimum average   cost is achieved, we have $H(\hat{Y})= H(\mathbf{Y})$. Thus, the codeword sequence approximates an i.i.d. sequence generated by distribution $\{\hat{P}_i\}$ (see Remark~\ref{remark::iid}).	\qed
	\end{remark}
	
	Given a prefix-free variable-length shaping code $\phi: \mathcal{X}^q \rightarrow \mathcal{Y}^*$, assume that after $nq$ source symbols are encoded, the codeword sequence is $\phi(x_1^{nq})$.   As in equations  (\ref{equ:averagewearcost1}),  (\ref{equ:averagewearcost2}),  (\ref{equ:averagewearcost3}), we formally define the expected  cost per source symbol, or \textit{total   cost} of a   shaping code as
	\begin{equation}
	\begin{split}
	T(\phi(\mathbf{X}^q))& = \frac{\sum_i E(N_i(\phi(X^{nq}))) C_i }{nq} = \frac{\sum_i E(N_i(\phi(X^q)))C_i}{q}\\
	&=\frac{E(L)}{q}\frac{\sum_{i} E(N_i(\phi(X^q)))C_i}{E(L)} = f\sum_{i=1}^v \hat{P_i}C_i.
	\end{split}
	\end{equation}
	
	We refer to the problem of  minimizing the total cost  as the \textit{type-\Romannum{2} shaping problem}. Shaping codes that achieve the minimum total  cost are referred to as \textit{optimal type-\Romannum{2} shaping codes}. The corresponding optimization problem is as follows:
	\begin{equation}
	\begin{aligned}
	& \underset{\hat{P}_i,f}{\text{minimize}}
	& & f\sum_{i=1}^v\hat{P}_iC_i\\
	& \text{subject to}
	& & H(\hat{Y})\geq H(\mathbf{Y}) = \frac{H(\mathbf{X})}{f}\\
	&  && \sum_{i}\hat{P}_i=1.
	\end{aligned}
	\end{equation} 	
	
\begin{figure}[h]
		\centering
		\includegraphics[width=1\columnwidth]{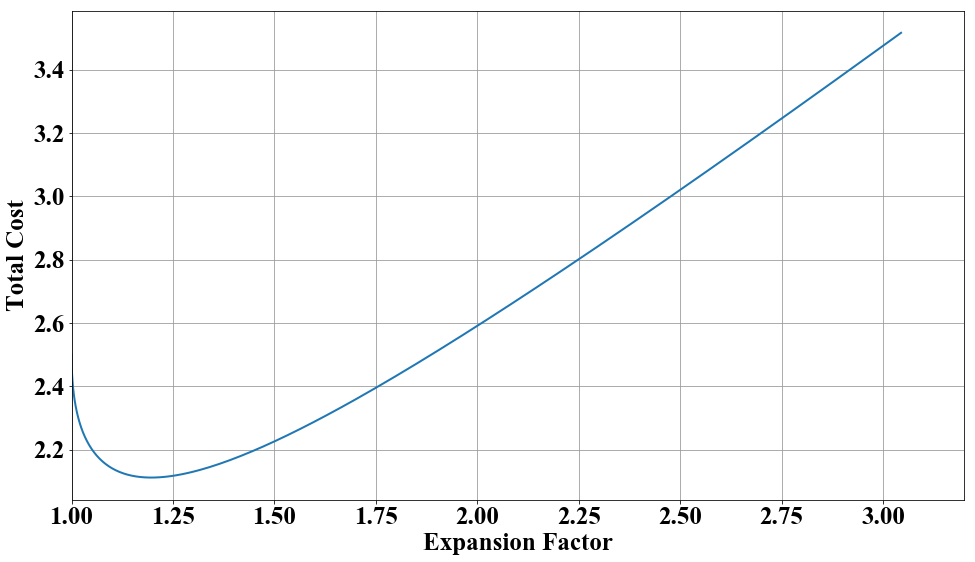}
		\caption{Total   cost versus $f$ for random source with $\mathcal{C}=[1,2,3,4]$. }
		\label{fig:costvsf}
	\end{figure}

	Using Theorem~\ref{performance:opt}, we can calculate the total   cost as a function of the expansion factor $f$. Fig.~\ref{fig:costvsf} shows the total   cost curve  for a quaternary source and code alphabet, a uniformly distributed source $\mathbf{X}$, and  cost vector $\mathcal{C}=[1,2,3,4]$.  There is an optimal value of $f$ and  corresponding minimum total   cost.
	
	We now determine the minimum achievable total   cost of a shaping code. 
	
	\begin{theorem}
		\label{opt_expansion}
		Given the source $\mathbf{X}$ and cost vector $\mathcal{C}$, if $C_1 \neq 0$, then the minimum total   cost of a shaping code $\phi: \mathcal{X}^q \rightarrow \mathcal{Y}^*$ is given by $f\sum_{i}\hat{P}_iC_i$, where $\hat{P}_i=2^{-\mu C_i}$, $\mu$ is a positive constant selected such that $\sum_i 2^{-\mu C_i}=1$. The corresponding expansion factor $f$ is
		\begin{equation}
		%\color{red}
		f =  \frac{H(\mathbf{X})}{-\sum_i \hat{P}_i \log_2 \hat{P}_i}.
		\end{equation}
		If $C_1 = 0$, then the total cost is a decreasing function of $f$. %\qed
	\end{theorem}
	\begin{IEEEproof}
		See Appendix~\ref{appen:2}.
	\end{IEEEproof}
	
	\begin{remark}
		\label{rmk::comparetotalcost}
		For a positive cost vector $\mathcal{C}$, the minimum achievable total  cost is 
		\begin{equation}
		\begin{split}
		&T(\phi(\mathbf{X}^q)) = f\sum_i \hat{P}_iC_i  =\frac{H(\mathbf{X})}{-\sum_i \hat{P}_i \log_2 \hat{P}_i}\sum_i \hat{P}_i C_i\\
		& =\frac{H(\mathbf{X})}{-\sum_i \hat{P}_i \log_2 2^{-\mu C_i}}\sum_i \hat{P}_i C_i = \frac{H(\mathbf{X})}{\mu\sum_i \hat{P}_i C_i}\sum_i \hat{P}_i C_i\\
		& = \frac{H(\mathbf{X})}{\mu}.
		\end{split}
		\end{equation}
		In~\cite[Theorem~4.4]{CsiszarKorner} and~\cite[Theorem~1]{HanUchida}, the minimum total cost of a prefix-free variable-length code was determined.  The capacity of a noiseless finite-state costly channel,  which is essentially the inverse of the minimum total cost, was considered in~\cite{McElieceRodemich},~\cite{Justesen},~\cite{Khandekar},~\cite{BochererSCC}, and~\cite{BochererThesis} from combinatorial and probabilistic perspectives.   Equivalences between the combinatorial and probabilistic definitions of capacity were established, extending the original results of Shannon.  
		However, these works did not address the code expansion factor and asymptotic symbol occurrence probability corresponding to the minimum total cost. 
		
		In~\cite[Problems 1.8]{McElieceBook},~\cite{Marcus} and~\cite[Sec. 5.2]{BochererThesis}, the relationship between the maximum entropy of a probability mass function on an alphabet with cost subject to an average cost constraint was discussed. However, these works did not explore the functional relationship between the total cost and the expansion factor of a code. Here, using the word-valued source perspective, we establish the relationship between the total cost of a rate-constrained prefix-free code and its expansion factor. This relationship plays an important role in the proof of the separation theorem  (Theorem~\ref{separation:type1}) in Appendix~\ref{appen:type1}. We also address the special case of zero lowest cost, i.e.,  $C_1 =0$, in which no global minimum can be reached. 
	\end{remark}
	
	\begin{remark}
		\label{remark:compression}
		If we only apply optimal lossless compression to the source $\mathbf{X}$, the code sequence has a uniform distribution. Therefore, we have $\mu = 0$ and $N=|\mathcal{Y}| > 1$. This implies that simply applying compression to the source data is not the best way to reduce the total   cost. \qed
	\end{remark}
	
	\subsection{Optimal Data Shaping Code Design}
	Many previous works investigated type-\Romannum{2} shaping code design. For example, see~ \cite{Karp}, ~\cite{Varn}, and~\cite{CsiszarKorner} . In this subsection, we consider the problem of designing an optimal type-\Romannum{1} shaping code by transforming this problem into a type-\Romannum{2} shaping problem. 
	Combining Theorems~\ref{performance:opt} and~\ref{opt_expansion}, we can prove the following equivalence theorem.
	
	\begin{theorem}
		\label{thm:typeonetypetwo}
		A code that achieves the minimum total cost for cost vector $\mathcal{C}'$ also achieves minimum average cost for cost vector $\mathcal{C}$ and expansion factor $f$ if
		\begin{equation}
		C_i' = -\log_2 \hat{P}_i,
		\end{equation}
		where $\{\hat{P}_i\}$ are the probabilities minimizing average cost for the cost vector $\mathcal{C}$ and expansion factor $f$. 
	\end{theorem}
	
	\begin{IEEEproof}
		First we consider the optimal type-\Romannum{2}  shaping code $\phi: \mathcal{X}^q\rightarrow \mathcal{Y}^*$ with cost vector $\mathcal{C}'$. By Theorem~\ref{opt_expansion}, this code generates codeword sequence with probability of occurrence $P_i' =2^{-\mu C_i'} $, where $\mu$ satisfies the equation 
		\begin{equation}
		\label{muresult}
		\sum_i 2^{-\mu C_i'} = 1.
		\end{equation}
		Since $C_i' = -\log_2 \hat{P}_i$, it is easy to check that the solution of equation (\ref{muresult}) is $\mu = 1$. This means when the minimum total   cost is achieved, the probability of occurrence of codeword sequence is $P_i' = 2^{-C_i'} = \hat{P}_i$ and the expansion factor of this code is
		\begin{equation}
		f' = \frac{H(X)}{-\sum_i P_i'\log_2 P_i'} = f
		\end{equation}
		Referring to Theorem~\ref{performance:opt}, we see that $\phi$ is also optimal with respect to minimizing average   cost with cost vector $\mathcal{C}$ and expansion factor $f$.
	\end{IEEEproof}
	
	When designing a type-\Romannum{1} shaping code with expansion factor $f$ and cost vector $\mathcal{C}$, we can first
	calculate the desired distribution $\{\hat{P}_i\}$, then transform this problem into a type-\Romannum{2}  shaping code problem for the channel with symbol cost  $\{C_i' = -\log_2 \hat{P}_i\}$. Thus we can apply known type-\Romannum{2} shaping code algorithms to solve this problem.

	\begin{remark}
		For an arbitrary i.i.d. source and a positive cost vector $\mathcal{C}$, generalized Shannon-Fano codes~\cite[Theorem 4.4]{CsiszarKorner} are tree-based variable-length codes, $\phi: \mathcal{X}^q\rightarrow \mathcal{Y}^*$, whose total cost is upper bounded by 
		\begin{equation}
		\begin{aligned}
		T(\phi) &< \frac{H(\mathbf{X})}{\mu} + \frac{\max_i \{C_i\}}{q}\\
		&\rightarrow \frac{H(\mathbf{X})}{\mu}\quad \text{as $q\rightarrow \infty$.}
		\end{aligned}
		\end{equation}
		This coding scheme includes dividing the $[0,1]$ interval based on $P(x_1^q)$ and calculating $2^{-\mu W}$, where $W$ is the cost of a codeword. This construction may become impractical when $q$ is large.
	\end{remark}

	\begin{remark}
		\label{Varn}
		For a uniform i.i.d. source and a positive cost vector $\mathcal{C}$, Varn codes \cite{Varn} are tree-based, variable-length codes $\phi_K: \mathcal{X}^{\log_{|\mathcal{X}|} K} \rightarrow \mathcal{Y}^*$ that minimize total cost for a specified codebook size $K$.  In \cite{Savari}, bounds were established on the average \textit{codeword} cost for the Varn code with codebook size $K$, denoted $C(\phi_K)$. Specifically, 
		\begin{equation}
		\frac{\log_2 K}{\mu} \leq C(\phi_K) \leq \frac{\log_2 K}{\mu} + \max_i \{C_i\}.
		\end{equation}
		Dividing by $\log_{|\mathcal{X}|}K$, we see that the total cost  of the Varn code with codebook size $K$ is bounded by
		\begin{equation}
		\frac{\log_2 |\mathcal{X}|}{\mu} \leq T(\phi_K) \leq \frac{\log_2 |\mathcal{X}|}{\mu} + \frac{\max_i \{C_i\}}{\log_{|\mathcal{X}|} K}.
		\end{equation}
		Therefore 
		\begin{equation} \lim_{K\rightarrow \infty}T(\phi_K) = \frac{\log_2 |\mathcal{X}|}{\mu}
		\end{equation}
		which implies that Varn codes are asymptotically optimal type-\Romannum{2} shaping codes  (see Remark~\ref{rmk::comparetotalcost}). \qed
	\end{remark}
	%%%%%%%%%%%%%%%%%%%%%%%%%%%%%%%%%%%%%%%%%%%%%%%%%
	%%%%%%%%%%%%%%%%%%%%%%%%%%%%%%%%%%%%%%%%%%%%%%%%%
	
	We now present a separation theorem  for  type-\Romannum{2} shaping codes. It states that minimum total cost can be achieved by a concatenation of optimal lossless compression with an optimal type-\Romannum{2} shaping code  for a uniform i.i.d. source.
	The proof uses a construction based on typical sequences. 
	\begin{theorem}
		\label{separation}
		Given the source $\mathbf{X}$ and cost vector $\mathcal{C}$, the minimum total   cost can be achieved by a concatenation of an optimal lossless compression code with an optimal type-\Romannum{2} shaping code for a uniform i.i.d. source.%\qed
	\end{theorem}
	\begin{IEEEproof}
		See Appendix~\ref{appen:3}.
	\end{IEEEproof}
	
	An example of an optimal type-\Romannum{2} shaping scheme that illustrates Theorem~\ref{separation} was described by Iwata in~\cite{Iwata}. It uses a concatenation of an LZ78 code and a Varn code as outer and inner codes, respectively.
	
	There is also a separation theorem  for   type-\Romannum{1}  shaping codes, stating that   minimum average cost for a given expansion factor  can be achieved by a concatenation of optimal lossless compression with an optimal type-\Romannum{1} shaping code  for a uniform i.i.d. source and suitable expansion factor.  The proof relies on the  type-\Romannum{2} separation theorem and the equivalence between type-\Romannum{2} and type-\Romannum{1} shaping codes established in Theorem~\ref{thm:typeonetypetwo}. It requires an analysis of the behavior of the total cost function in the vicinity of the expansion factor that minimizes total cost.
	
	\begin{theorem}
		\label{separation:type1}
		Given the source $\mathbf{X}$, cost vector $\mathcal{C}$ and expansion factor $f$, the minimum average   cost can be achieved by a concatenation of an optimal lossless compression code with a binary optimal type-\Romannum{1} shaping code for uniform i.i.d. source and expansion factor
		\begin{equation}
		f' = \frac{f}{H(\mathbf{X})}.
		\end{equation}%\qed
	\end{theorem}
	\begin{IEEEproof}
		See Appendix~\ref{appen:type1}.
	\end{IEEEproof}

	%%%%%%%%%%%%%%%%%%%%%%%%%%%%%%%%%%%%%%%%%%%%%%%%%
	%%%%%%%%%%%%%%%%%%%%%%%%%%%%%%%%%%%%%%%%%%%%%%%%%
	
	\section{Distribution Matching Code Design}
	\label{sec:DM}
	Given a target distribution $\{P_i\}$, distribution matching (DM) considers the problem of mapping an i.i.d. sequence of source symbols to an output sequence of  symbols that are approximately independent and distributed according to $\{P_i\}$. An optimal DM code  must satisfy two conditions: the codeword sequence has symbol occurrence probabilities $\hat{P}_i = P_i$, and the output sequence looks like an i.i.d. sequence. We measure the latter property using the asymptotic normalized  KL-divergence defined in Lemma~\ref{upperbound_marginaldistribution}.
	
	It has been shown in Theorems~\ref{performance:opt} and \ref{opt_expansion} that an optimal shaping code will generate an output sequence such that $\lim_{l\rightarrow \infty} \frac{1}{l}D(Y_1^l || \hat{Y}_1^l) = 0$. Thus the output sequence sequence approximates an i.i.d. sequence with symbol occurrence probability distribution $\{\hat{P_i}\}$. This implies that we can solve the distribution matching problem by designing a corresponding shaping code. In this section, we consider the problem of designing optimal DM codes. We first formulate the problem of \textit{generating an i.i.d. sequence} and then show the connection between DM codes and  shaping codes. We then propose a \textit{generalized expansion factor} to measure the performance of a DM code. A comparison of DM code performance measures is also presented.

	%%%%%%%%%%%%%%%%%%%%%%%%%%%%%%%%%%%%%%%%%%%%%%%%%
	%%%%%%%%%%%%%%%%%%%%%%%%%%%%%%%%%%%%%%%%%%%%%%%%%
	\subsection{Problem Formulation}
	We use the asymptotic normalized Kullback-Leibler divergence~\cite{Soriaga} to formally define an optimal DM code $\phi$ for distribution $\{P_i\}$.
	\begin{definition}
		A variable-length mapping  $\phi: \mathcal{X}^q\rightarrow \mathcal{Y}^*$ is an optimal DM code for distribution  $\{P_i\}$ if 
		\begin{equation}
		\lim_{l\rightarrow \infty} \frac{1}{l}D(Y_1^l || \tilde{Y}_1^l) = 0,
		\end{equation}
		where $\mathbf{\tilde{Y}}$ is an i.i.d. process with distribution $\{P_i\}$.\qed
	\end{definition}
	
	By combining Theorem~\ref{Hiroyoshitheorem} and Lemma~\ref{upperbound_marginaldistribution}, we can prove the following theorem.
	\begin{theorem}
		\label{thm:expanofdistribuion}
		The expansion factor of a mapping satisfies the lower bound
		\begin{equation}
		f = \frac{H(\mathbf{X})}{H(\mathbf{Y})} \geq \frac{H(\mathbf{X})}{H(\hat{Y})}
		\end{equation}
		with equality if and only if $\lim_{l\rightarrow \infty} \frac{1}{l}D(Y_1^l || \hat{Y}_1^l) = H(\hat{Y}) - H(\mathbf{Y}) = 0$. When $f = \frac{H(\mathbf{X})}{H(\hat{Y})}$, this code is an optimal DM code for distribution $\{{P_i}\}$. \qed
	\end{theorem}
	\begin{remark}
		\label{randomiid}
		Assuming this mapping is an optimal compression, the compression ratio $g$ is
		\begin{equation}
		g = \frac{H(\mathbf{X})}{\log_2 |\mathcal{Y}|}.
		\end{equation}
		By Theorem~\ref{Hiroyoshitheorem} and Lemma~\ref{upperbound_marginaldistribution}, we have
		\begin{equation}
		H(\hat{Y}) \geq H(\mathbf{Y}) = \frac{H(\mathbf{X})}{g} = \log_2 |\mathcal{Y}|.
		\end{equation}
		Since $H(\hat{Y}) \leq \log_2 |\mathcal{Y}|$, we know that $H(\hat{Y}) = H(\mathbf{Y}) = \log_2 |\mathcal{Y}|$ and 
		\begin{equation}
		\lim_{l\rightarrow\infty} \frac{1}{l}D(Y_1^l||\hat{Y}_1^l) = 0.
		\end{equation}
		This implies the codeword sequence looks i.i.d. and has probability of occurrence
		\begin{equation}
		\hat{P_i} = \frac{1}{|\mathcal{Y}|}\quad \text{for all $i$}.
		\end{equation}
		This proves the well-known fact that the output of an optimal compression approximates a uniform i.i.d. sequence \cite{Vis1998},\cite{HanBook},\cite{HanUchida}.\qed
	\end{remark}
	%%%%%%%%%%%%%%%%%%%%%%%%%%%%%%
	Let $\mathbf{\tilde{Y}}$ be the i.i.d. process with distribution $\{P_i\}$.  As in the derivation of (\ref{I_divergence3}) in Lemma~\ref{upperbound_marginaldistribution}, we find
	\begin{equation}
	\label{equ::kltypeone}
	\begin{aligned}
	&\lim_{l\rightarrow \infty} \frac{1}{l} D(Y_1^l || \tilde{Y}_1^l) \\ &=-\lim_{l\rightarrow \infty} \frac{1}{l}H(Y_1^l)  - \sum_i \log_2 P_i\lim_{l\rightarrow \infty}\sum_{y_1^l\in \mathcal{Y}^l} \frac{Q(y_1^l) {N_i(y_1^l)}}{l}\\
	&= -H(\mathbf{Y}) - \sum_i \hat{P_i} \log_2 P_i = -\sum_{i} \hat{P_i} \log_2 P_i - \frac{H(\mathbf{X})}{f}\\
	& = \frac{-f\sum_{i} \hat{P_i} \log_2 P_i - H(\mathbf{X})}{f}.
	\end{aligned}
	\end{equation}
	%%%%%%%%%%%%%%%%%%%%%%%%%%%%%%
	
	From Theorem~\ref{opt_expansion}, we know that for a channel with cost $\{C_i = -\log_2 P_i\}$, the total cost $-f\sum_{i} \hat{P_i} \log_2 P_i$ is lower bounded by $H(\mathbf{X})$. The shaping code that achieves this lower bound has the following two properties:
	\begin{itemize}
		\item The probability of occurrence of symbol $\beta_i$ satisfies $\hat{P}_i = P_i$ for all $\beta_i$,
		\item The asymptotic normalized  KL-divergence between $\mathbf{Y}$ and $\hat{Y}$ satisfies
		\begin{equation}
		\lim_{l\rightarrow \infty}\frac{1}{l} D(Y_1^l|| \hat{Y}_1^l) = 0.
		\end{equation}
	\end{itemize}
	This implies that this code generates a sequence that approximates an i.i.d. sequence with distribution $\{P_i\}$. 
	This analysis also implies that the expansion factor of an optimal DM code is
	\begin{equation}
	f_\text{opt} = \frac{H(\mathbf{X})}{-\sum_i \hat{P_i}\log_2 \hat{P_i}}= \frac{H(\mathbf{X})}{-\sum_i P_i\log_2 P_i}.
	\end{equation}
	
	We summarize in the following theorem the relationship between optimal shaping codes and optimal DM codes, extending the result in~\cite{HanUchida} by explicitly showing the optimal expansion factor.
	\begin{theorem}
		\label{thm:shapinganddm}
		The optimal type-\Romannum{2} shaping code with cost vector $\mathcal{C}$, or the  equivalent type-\Romannum{1} shaping code  from Theorem~\ref{thm:typeonetypetwo},  is an optimal DM code for distribution $\{P_i\}$ if 
		\begin{equation}
		\label{equ::shell}
		C_i = -\log_2 P_i
		\end{equation}
		for every symbol $\beta_i$, in the sense that
		\begin{equation}
		\lim_{l\rightarrow \infty} \frac{1}{l}D(Y_1^l || \tilde{Y}_1^l) = 0.
		\end{equation} The expansion factor of this optimal DM code is
		\begin{equation}
		f_\text{opt} =  \frac{H(\mathbf{X})}{-\sum_i P_i\log_2 P_i}.
		\end{equation}\qed
	\end{theorem}

	\begin{remark}
		Shell mapping was used in~\cite{SchulteSteiner}   to design fixed-length DM codes with uniformly distributed input bits. The shell mapper that minimizes  informational divergence (introduced later in Section~\ref{sec::comparegef}) uses the ``self-information'' weight function $C_i = -\log_2 P_i$ and the optimal expansion factor is determined by a search. Theorem~\ref{thm:shapinganddm} considers a more general variable-length DM code with arbitrary i.i.d. source and characterizes the optimal expansion factor. Codes minimizing informational divergence are discussed further in Section~\ref{sec::comparegef}.
	\end{remark}
	
	%%%%%%%%%%%%%%%%%%%%%%%%%%%%%%%%%%%%%%%%%%%%%%%%%
	%%%%%%%%%%%%%%%%%%%%%%%%%%%%%%%%%%%%%%%%%%%%%%%%%
	\section{Generalized Expansion Factor}
	\label{sec:gef}
	%%%%%%%%%%%%%%%%%%%%%%%%%%%%%%%%%%%%%%%%%%%%%%%%%
	%%%%%%%%%%%%%%%%%%%%%%%%%%%%%%%%%%%%%%%%%%%%%%%%%
	The relationship between optimal shaping codes and optimal DM codes was established above. The total   cost of the shaping code suggests an alternative performance measure for a DM code which will be useful when analyzing the optimality of a shaping-based DM code construction and in proving a separation theorem for DM codes. Specifically, we define the \textit{generalized expansion factor} (GEF)  of a prefix-free variable-length code as follows.
	\begin{definition}
		Given a prefix-free variable-length code $\phi: \mathcal{X}^{q}\rightarrow \mathcal{Y}^*$ and a set of positive real numbers $\{P_1,P_2,\ldots,P_v\}$ such that $\sum P_i =1$,
		the {\bf generalized expansion factor} of this code is defined as
		\begin{equation}
		\label{func::gefdef}
		F(\phi,P_1,\ldots,P_v) =- f\frac{\sum_i \hat{P_i}\log_2 P_i}{\log_2 |\mathcal{Y}|}
		\end{equation}
		where $f$ is the code expansion factor and $\{\hat{P_i}\}$ is the asymptotic symbol occurrence probability distribution.\qed
		%\vspace{-4ex}
	\end{definition}
	For simplicity, we sometimes use $F$ to represent $F(\phi,P_1,\ldots,P_v)$. The following theorem shows that $F$ can be used to evaluate an optimal DM code.
	\vspace{-2ex}
	\begin{theorem}
		\label{thm::geflowerbound}
		Given a prefix-free variable-length code $\phi: \mathcal{X}^{q}\rightarrow \mathcal{Y}^*$ and a set of positive real numbers $\{P_1,P_2,\ldots,P_v\}$ such that $\sum P_i =1$, the generalized expansion factor of this mapping is lower bounded by
		\vspace{-2ex}
		\begin{equation}
		F(\phi,P_1,\ldots,P_v) \geq \frac{H(\mathbf{X})}{\log_2 |\mathcal{Y}|}.
		\vspace{-1ex}
		\end{equation}
		If $F = \frac{H(\mathbf{X})}{\log_2 |\mathcal{Y}|}$, this mapping is an optimal DM code for the target distribution $\{P_i\}$, in the sense that
		\begin{equation}
		\lim_{l\rightarrow \infty} \frac{1}{l}D(Y_1^l || \tilde{Y}_1^l) = 0.
		\end{equation} \qed%~\qed
	\end{theorem}
	
	\begin{IEEEproof}
		Assume symbol $\beta_i$ in the codeword sequence has cost $C_i = -\log_2 P_i$. The total   cost of this mapping is
		\begin{equation}
		\label{func::totalcostvsgef}
		T(\phi) = f \sum_i \hat{P_i} C_i  = -f\sum_i \hat{P_i}\log_2 P_i.
		\end{equation}
		
		Comparing equations~(\ref{func::gefdef}) and~(\ref{func::totalcostvsgef}) we have 
		\begin{equation}
		\label{equ::totalgefequivalence}
		F(\phi,P_1,\ldots,P_v)  = \frac{T(\phi)}{\log_{2}|\mathcal{Y}|}.
		\end{equation}
		This indicates that the GEF of a DM code is equivalent to its total cost when applying it to a costly channel with cost $C_i = -\log_2 P_i.$ From Theorem~\ref{opt_expansion}, we know the total cost of a prefix-free mapping satisfies the lower bound
		\begin{equation}
		\label{equ::NEF}
		T(\phi) \geq \frac{H(\mathbf{X})}{\mu}
		\end{equation}
		where $\mu$ is a constant such that $\sum 2^{-\mu C_i} = 1$. Since $C_i = -\log_2 P_i$, it is easy to check that $\mu = 1$ and 
		\begin{equation}
		\label{equ::totalcostgefequivalence}
		F(\phi,P_1,\ldots,P_v)  = \frac{T(\phi)}{\log_{2}|\mathcal{Y}|} \geq \frac{H(\mathbf{X})}{\log_2 |\mathcal{Y}|}.
		\end{equation}
		When the minimum GEF is achieved, this code is also an optimal type-\Romannum{2} shaping code with $C_i = -\log_2 P_i$. Theorem~\ref{thm:shapinganddm} then implies that this code is an optimal DM code for the target distribution $\{P_i\}$, in the sense that
		\begin{equation}
		\lim_{l\rightarrow \infty} \frac{1}{l}D(Y_1^l || \tilde{Y}_1^l) = 0.
		\end{equation} 
	\end{IEEEproof}
	
	\begin{remark}
		\label{remark:varnasdm}
		As shown in Remark~\ref{Varn}, for a uniform i.i.d. source and a cost vector $\mathcal{C}$, a Varn code $\phi_K: \mathcal{X}^{\log_{|\mathcal{X}|} K} \rightarrow \mathcal{Y}^*$ is an asymptotically optimal type-\Romannum{2} shaping code. 
		%In \cite{Savari}, the author proved a tighter upper bound on its total   cost. 
		If the costs are given by $C_i= -\log_2 P_i$, where $\sum_i P_i = 1$, the total   cost is bounded by
		\begin{equation}
		\log_2 |\mathcal{X}| \leq T(\phi_K) \leq \log_2 |\mathcal{X}| + \frac{\max_i \{C_i\}}{\log_{|\mathcal{X}|} K}.
		\end{equation}
		
		Equation~(\ref{equ::totalgefequivalence}) implies that for the target distribution $\{P_i\}$, Varn codes minimize GEF for a specified codebook size $K$. Thus, a Varn code can be regarded as a DM code with generalized expansion factor bounded by
		\begin{equation}
		\frac{\log_2 |\mathcal{X}|}{\log_2 |\mathcal{Y}|}\leq F\leq \frac{\log_2 |\mathcal{X}|}{\log_2 |\mathcal{Y}|}(1 + \frac{\max_i\{-\log_2 P_i\}}{\log_{2} K}).
		\end{equation}
		Therefore, we have 
		\begin{equation} \lim_{K\rightarrow \infty}F(\phi_K,P_1,\ldots,P_v) = \frac{\log_2 |\mathcal{X}|}{\log_2 |\mathcal{Y}|}
		\end{equation}
		which implies that Varn codes are asymptotically optimal DM codes. Fig.~\ref{Tunstall23_distribution} and Fig.~\ref{Tunstall23_normalized_rate} show the probability of occurrence and generalized expansion factor of binary Varn codes (i.e.,  with $\mathcal{X} = \mathcal{Y} = \{0,1\}$) for a target distribution $P_0=2/3$, $P_1 = 1/3$. As the codebook size $K$ increases, we see that the probability of occurrence $\hat{P_0}$ approaches the target distribution value $P_0 =2/3$ and the generalized expansion factor  approaches the theoretical lower bound $1$.\qed
		\begin{figure}
			\centering
			\includegraphics[width=1\columnwidth]{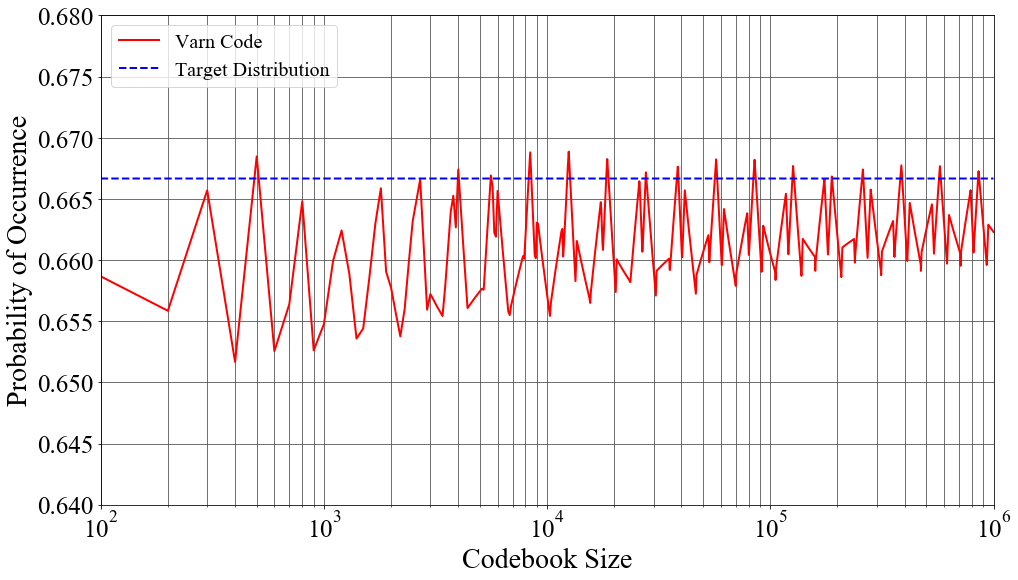}
			\caption{Probability of occurrence $\hat{P_0}$ of a Varn code for the target distribution $\{2/3, 1/3\}$.}
			\label{Tunstall23_distribution}
		\end{figure}
		
		\begin{figure}
			\centering
			\includegraphics[width=1\columnwidth]{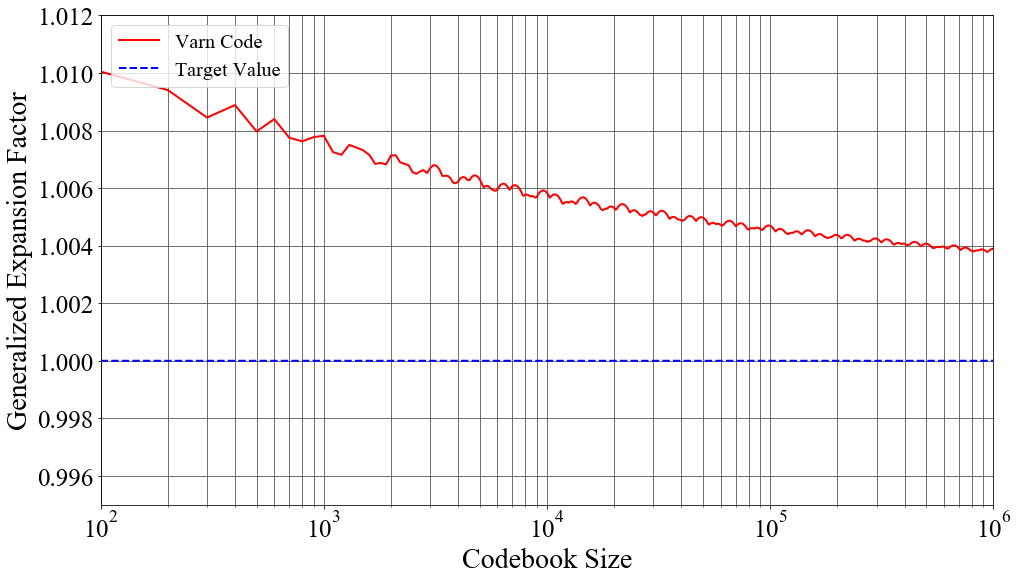}
			\caption{Generalized expansion factor of a Varn code for the target distribution $\{2/3, 1/3\}$.}
			\label{Tunstall23_normalized_rate}
		\end{figure}
	\end{remark}
	
	The separation theorem for shaping codes in Theorem~\ref{separation}  now extends naturally to DM codes.
	\begin{theorem}
		\label{separation:DM}
		An optimal DM code can be constructed by a concatenation of optimal lossless compression with an optimal DM code for a uniform i.i.d. source, in the sense that the minimum generalized expansion factor can be achieved by such a concatenation. \qed
	\end{theorem}
	
	\begin{remark}
		When $P_1 = P_2 = \cdots = P_v = \frac{1}{|\mathcal{Y}|}$, the generalized expansion factor reduces to 
		\begin{equation}
		F = f \frac{\sum_i \hat{P_i} \log_2 |\mathcal{Y}|}{\log_2 |\mathcal{Y}|} = f.
		\end{equation}
		This provides the motivation for designating $F$ by this name.\qed
	\end{remark}
	
	\begin{remark}
		\label{rmk::han}
		We use an example to illustrate  the difference between the generalized expansion factor and the normalized conditional divergence introduced in~\cite{HanUchida} and~\cite{HanBook} when the encoder has finite length.
		Given a ternary source with alphabet $\mathcal{X}=\{\alpha_1,\alpha_2,\alpha_3\}$ and probability distribution $\{\frac{1}{2}, \frac{1}{4},\frac{1}{4}\}$,  consider two  codes defined by the mappings 
		\begin{equation}
		\begin{aligned}
		&\Phi_1: \{\alpha_1 \rightarrow 0, \alpha_2 \rightarrow 10, \alpha_3 \rightarrow 11  \},\\
		&\Phi_2: \{\alpha_1 \rightarrow 00, \alpha_2 \rightarrow 10, \alpha_3 \rightarrow 11  \}.
		\end{aligned}
		\end{equation}
		Their generalized expansion factors for target distribution $\{1/2,1/2\}$ are 
		\begin{equation}
		F_1 = \frac{3}{2} <  F_2 = 2.
		\end{equation}
		This suggests that $\Phi_1$ is a better approximation of an optimal DM code for target distribution $\{\frac{1}{2}, \frac{1}{2}\}$
		(in fact, $\Phi_1$ is an optimal DM code).	
		The normalized conditional divergences are
		\begin{equation}
		\begin{aligned}
		D&(\Phi_1(X)||V|I)=\\ &\frac{1}{2}(\log_2 \frac{1}{1/2})  +\frac{1}{2}(\frac{1}{2}\log_2 \frac{1/2}{1/4} + \frac{1}{2}\log_2 \frac{1/2}{1/4}) = 1
		\end{aligned}
		\end{equation}
		\begin{equation}
		\begin{aligned}
		D&(\Phi_2(X)||V|I) \\&= \frac{1}{2}\log_2 \frac{1/2}{1/4} + \frac{1}{4}\log_2 \frac{1/4}{1/4} + \frac{1}{4}\log_2 \frac{1/4}{1/4} = \frac{1}{2}.
		\end{aligned}
		\end{equation}
		We find that $D(\Phi_1(X)||V|I) > D(\Phi_2(X)||V|I)$, which suggests the opposite conclusion that $\Phi_2$ would be a better approximation of the optimal DM code.\qed
	\end{remark}
	%\vspace{-5ex}
	%%%%%%%%%%%%%%%%%%%%%%%%%%%%%%%%%%%%%%%%%%%%%%%%%%%%%%%%%%%%%%%%%%%%%%%%%%%%%%%%%%%%%%%%%%%%%%%%%%
	\section{Comparison of DM Performance Measures}
	\label{sec::compare}
	In this section, we use a shaping code perspective to study DM codes whose performance is measured using informational divergence and normalized informational divergence. 
	\subsection{Generalized Expansion Factor and Informational Divergence}
	\label{sec::comparegef}
	In this subsection we study the relationship between the generalized expansion factor and the informational divergence introduced in \cite{AmjadBocherer}, which is also used as a performance measure for DM codes. 
	
	Consider a variable-length code $\phi: \mathcal{X}^{\log_{|\mathcal{X}|} K} \rightarrow \mathcal{Y}^*$  with codebook size $K$. We use $\mathcal{L}$ to denote the set of all codewords generated by this mapping. The leaf probability, or the probability of a codeword $y_1^l$, is defined as
	\begin{equation}
	P^{\mathcal{L}}(y_1^l) = P(y_1)P(y_2)\ldots P(y_l) = \prod P_i ^{N_i(y_1^l)}.
	\end{equation}
	This is also the probability of sequence $y_1^l$ generated by an i.i.d. source with distribution $\{P_i\}$. The true probability of codeword $y_1^l$ is the probability of the corresponding source sequence $\phi^{-1}(y_1^l)$. The \textit{informational divergence} ({I-divergence}) between these two distributions is defined as
	\begin{equation}
	I = \sum_{y_1^l \in \mathcal{L}} P(\phi^{-1}(y_1^l)) \log_2 \frac{P(\phi^{-1}(y_1^l))}{P^{\mathcal{L}}(y_1^l)}.
	\end{equation}
	Now we use the same code for type-\Romannum{2}  shaping. We set the cost of each symbol to be $C_i = -\log_2 P_i$. The cost of codeword $y_1^l$ is 
	\begin{equation}
	\begin{aligned}
	W(y_1^l)& = \sum C_i N_i(y_1^l)  = -\sum \log_2 P_i ^{N_i(y_1^l)}\\
	& = -\log_2 \prod P_i^{N_i(y_1^l)} = -\log_2 P^{\mathcal{L}}(y_1^l)
	\end{aligned}
	\end{equation}
	and the total   cost of this shaping code, or equivalently the GEF, is
	\begin{equation}
	\begin{aligned}
	F&(\phi,P_1,\ldots,P_v) = \frac{T(\phi)}{\log_2 |\mathcal{Y}|}\\& = \frac{1}{\log_{|\mathcal{X}|} K \log_2 |\mathcal{Y}|}\sum_{y_1^l \in \mathcal{L}} P(\phi^{-1}(y_1^l))W(y_1^l)\\
	&= -\frac{\log_2 |\mathcal{X}|}{\log_{2} K\log_2 |\mathcal{Y}|}\sum_{y_1^l \in \mathcal{L}} P(\phi^{-1}(y_1^l)) \log_2 P^{\mathcal{L}}(y_1^l).
	\end{aligned}
	\end{equation}
	The {I-divergence} of this code can then be expressed in terms of its GEF, namely
	\begin{equation}
	\label{equ:IDivergencetocost}
	\begin{aligned}
	I& = \sum_{y_1^l \in \mathcal{L}} P(\phi^{-1}(y_1^l)) \log_2 \frac{P(\phi^{-1}(y_1^l))}{P^{\mathcal{L}}(y_1^l)}\\
	& =\sum_{y_1^l \in \mathcal{L}}  P(\phi^{-1}(y_1^l)) \log_2  P(\phi^{-1}(y_1^l))\\&-  \sum_{y_1^l \in \mathcal{L}}P(\phi^{-1}(y_1^l)) \log_2 P^{\mathcal{L}}(y_1^l)\\
	& = F\frac{\log_{2}K \log_2 |\mathcal{Y}|}{\log_2 |\mathcal{X}|} - H(\mathbf{X}^{\log_{|\mathcal{X}|}  K}) \\
	& =( F - \frac{H(\mathbf{X})}{\log_2 |\mathcal{Y}|})\frac{\log_{2}K \log_2 |\mathcal{Y}|}{\log_2 |\mathcal{X}|}.
	\end{aligned}
	\end{equation}
	Since $\log_2 K$ is a constant, minimizing $I$ is equivalent to minimizing $F$. This equation shows the relationship between {I-divergence} and GEF, and also highlights the duality between costly channel coding and DM coding. As shown in Remark~\ref{remark:varnasdm}, Varn codes minimize GEF for a uniform i.i.d. source. Therefore we can conclude the following optimality theorem for Varn codes.
	\begin{theorem}
		\label{thm::varnminimizeidivergence}
		Let $\{P_i\}$ be a target distribution. A code $\phi: \mathcal{X}^{\log_{|\mathcal{X}|} K}\rightarrow \mathcal{Y}^*$ for a uniform i.i.d. source that minimizes {I-divergence} is given by a Varn code designed for costs  $C_i = -\log_2 P_i$.\qed
		%\vspace{-3ex}
	\end{theorem}
	\begin{remark}
		In~\cite{Savari}, Savari showed that Varn codes and reverse Tunstall codes are identical when finding exhaustive prefix-free codes (i.e., when $(K-1)/(|\mathcal{Y}| - 1)$ is an integer).  Specifically, a Tunstall code designed to compress distribution $\{P_i\}$ and a Varn code designed for costly channel $\{C_i = -\log_2 P_i\}$ generate identical code trees. Therefore a reverse Tunstall code minimizes the {I-divergence} when the target distribution is binary (i.e., when $|\mathcal{Y}|   = 2$). This was also proved in~\cite[Proposition 1]{AmjadBocherer} using a different method.
		
		However, for non-exhaustive codes this equivalence does not exist, and it remains unknown whether a reverse Tunstall code minimizes {I-divergence} when the target distribution is non-binary (i.e., when $|\mathcal{Y}| > 2$). Therefore Theorem~\ref{thm::varnminimizeidivergence} can be viewed as a generalization of~\cite[Proposition 1]{AmjadBocherer}.
	\end{remark}

	\subsection{Type-I Shaping Problem and Normalized I-Divergence}
	\label{subsec::tonenid}
	Another measure for DM codes used in~\cite{AmjadBocherer} is normalized {I-divergence}. In this subsection, we study its properties using the perspective of the type-\Romannum{1} shaping problem. Normalized {I-divergence} is defined as 
	\begin{equation}
	\mathscr{I}= \frac{I}{E(L)}.
	\end{equation}
	Using (\ref{equ:IDivergencetocost}), we rewrite this as
	\begin{equation}
	\label{equ::inorm}
	\begin{aligned}
	\mathscr{I} &= ( F - \frac{H(\mathbf{X})}{\log_2 |\mathcal{Y}|})\frac{\log_{2}K \log_2 |\mathcal{Y}|}{E(L)\log_2 |\mathcal{X}|} \\
	& = ( - f\frac{\sum_i \hat{P_i}\log_2 P_i}{\log_2 |\mathcal{Y}|} - \frac{H(\mathbf{X})}{\log_2 |\mathcal{Y}|})\frac{\log_2 |\mathcal{Y}|}{f}\\
	&= \sum_i \hat{P_i}C_i - \frac{H(\mathbf{X})}{f} %= \sum_i \hat{P_i}C_i - H(\mathbf{Y})
	\end{aligned}
	\end{equation}
	where $C_i = -\log_2 P_i$. From equations~(\ref{equ::kltypeone}) and~(\ref{equ::inorm}) we see that asymptotic normalized KL-divergence and normalized {I-divergence} are identical for i.i.d. distribution matching. 
	
	We divide the problem of finding the minimum $\mathscr{I}$ into two parts. First we fix the expansion factor $f$ and find the minimum achievable $\mathscr{I}$, denoted by $\mathscr{I}_\text{min}(f)$. Then we find the optimal $f$ to minimize $\mathscr{I}_\text{min}(f)$. 
	%Define $\mathscr{I}_\text{min}(f)$ the minimum normalized {I-divergence} with a specified expansion factor $f$.  
	The result is found by noting the similarity to the type-\Romannum{1} shaping problem and invoking Theorem~\ref{performance:opt}.
	
	\begin{theorem}
		\label{thm::nidtypeone}
		Let $\phi$ be a prefix-free variable-length mapping with expansion factor $f$. Let $\{P_i\}$ be the target distribution and set $C_i = -\log_2 P_i$.
		The minimum normalized {I-divergence} $\mathscr{I}_\text{min}(f)$ with fixed $f$ is 
		\begin{equation}
		\label{equ::iminfthm}
		\mathscr{I}_\text{min}(f) = \sum_i\hat{P_i}C_i - \frac{H(\mathbf{X})}{f},
		\end{equation}
		where $\hat{P_i} = \frac{2^{-\mu C_i}}{\sum_i 2^{-\mu C_i}}$ and $H(\hat{Y}) = -\sum_{i} \hat{P_i}\log_2 \hat{P_i} = H(\mathbf{X}) / f$.
	\end{theorem}
	
	\begin{IEEEproof}
		We must solve the following optimization problem, which is closely related to the  type-\Romannum{1} shaping problem.
		\begin{equation}\label{}
		\begin{aligned}
		& \underset{\hat{P}_i}{\text{minimize}}
		& & \sum_{i}\hat{P}_iC_i - \frac{H(\mathbf{X})}{f}\\
		& \text{subject to}
		& &H(\hat{Y})\geq \frac{H(\mathbf{X})}{f}\\
		&  && \sum_{i}\hat{P}_i=1.
		\end{aligned}
		\end{equation}
		From Theorem~\ref{performance:opt}, we immediately have
		\begin{equation}
		\label{equ::iminandd}
		\mathscr{I}_\text{min}(f) = \sum_i \hat{P_i}C_i - \frac{H(\mathbf{X})}{f} = \sum_i \hat{P_i}C_i - H(\hat{Y}),
		\end{equation}
		where $\hat{P_i} = \frac{2^{-\mu C_i}}{\sum_j 2^{-\mu C_j}}$ and $H(\hat{Y}) = -\sum_{i} \hat{P_i}\log_2 \hat{P_i} = H(\mathbf{X}) / f$.
	\end{IEEEproof}
	The next proposition determines the derivative of $\mathscr{I}_\text{min}(f)$ and finds the optimal expansion factor, $f_\text{opt}$, that minimizes  $\mathscr{I}_\text{min}(f)$.
	\begin{prop}
		\label{prop::propertyiminf}
		The first derivative of $\mathscr{I}_\text{min}(f)$ is
		\begin{equation}
		\label{equ:didmu1}
		\frac{\mathrm{d}\mathscr{I}_\text{min}}{\mathrm{d}f} = \frac{H(\mathbf{X})}{f^2} \frac{\mu - 1}{\mu}\quad \mu > 0.
		\end{equation}
		Let $f_\text{opt}= -H(\mathbf{X}) /{\sum_i P_i \log_2 P_i}$. Then 
		$\mathscr{I}_\text{min}(f)$ is continuous, strictly monotone decreasing on $[\frac{H(\mathbf{X})}{\log_2 |\mathcal{Y}|}, f_\text{opt})$ (or, for $\mu\in [0,1)$) and continuous, strictly monotone increasing on $(f_\text{opt}, +\infty)$ (or, for $\mu\in (1,\infty)$). When $f = f_\text{opt}$, $\mathscr{I}_\text{min}(f_\text{opt}) =0$.
	\end{prop}
	\begin{IEEEproof}
		We have studied the behavior of minimum total cost with fixed $f$ in Appendices~\ref{appen:2} and~\ref{appen:type1}. Here we use the same technique to study $\mathscr{I}_\text{min}(f)$. Note that $\mathscr{I}_\text{min}(f)$ is a function of $\mu$. The derivative $\mathrm{d}f/\mathrm{d}\mu$ is already given in equation (\ref{equ:dfdmu}), and it is easy to check that 
		\begin{equation}
		\label{equ:didmu}
		\frac{\mathrm{d}\mathscr{I}_\text{min}}{\mathrm{d}\mu} = \frac{ (\mu - 1)\ln 2  \sum_{i< j}2^{-\mu (C_i+C_j)} (C_i - C_j )^2}{N^2}.
		\end{equation}
		Applying the chain rule along with (\ref{equ:didmu}) and (\ref{equ:dfdmu}), we have
		\begin{equation}
		\label{equ:didf}
		\frac{\mathrm{d}\mathscr{I}_\text{min}}{\mathrm{d}f} = \frac{H(\mathbf{X})}{f^2} \frac{\mu - 1}{\mu}.
		\end{equation}
		Let $f_\text{opt}=\frac{H(\mathbf{X})}{-\sum_i P_i \log_2 P_i}$ (or, equivalently, let $\mu = 1$). Equations~(\ref{equ:didmu}) and~(\ref{equ:didf}) imply that $\mathscr{I}_\text{min}(f)$ is continuous,  strictly monotone decreasing on $[\frac{H(\mathbf{X})}{\log_2 |\mathcal{Y}|}, f_\text{opt})$ (or, for $\mu\in [0,1)$) and strictly monotone increasing on $(f_\text{opt}, +\infty)$ (or, for $\mu\in (1,\infty)$). The minimum of $\mathscr{I}_\text{min}(f)$ is achieved when $f = f_\text{opt}$. We have
		\begin{equation}
		\begin{aligned}
		\mathscr{I}_\text{min}(f_\text{opt}) &= \sum_i \hat{P_i}C_i - \frac{H(\mathbf{X})}{f_{\text{opt}}} \\
		& = \sum_i \frac{2^{-C_i}}{\sum_j 2^{-C_j}} C_i+ \sum_i P_i \log_2 P_i \\
		& =-\sum_i \frac{P_i}{\sum_j P_j} \log_2 P_i+ \sum_i P_i \log_2 P_i  = 0.
		\end{aligned}
		\end{equation}
		This completes the proof.
	\end{IEEEproof}
	
	\begin{remark}
		In~\cite[Sec 5.1]{BochererThesis}, the author studied the minimum KL-divergence between a pmf  $\{\hat{P_i}\}$ and the target distribution $\{P_i\}$, where each $\hat{P_i}$ is associated with a cost $w_i$ and the average cost of the pmf is upper bounded by $C$. The analysis is similar to the analysis in Proposition~\ref{prop::propertyiminf} if we specialize to the case where $w_i = -\log_2 P_i = C_i$. The KL-divergence is
		\begin{equation}
		\begin{aligned}
		\label{equ::inormpmf}
		D &= D(\hat{P_i}|| P_i) = \sum_i \hat{P_i} \log_2 \frac{\hat{P_i}}{P_i} \\
		& = - \sum_i \hat{P_i} \log_2 P_i + \sum_i\hat{P_i} \log_2 \hat{P_i}\\
		& = \sum_i \hat{P_i} C_i- H(\hat{P}).
		\end{aligned}
		\end{equation} 
		By combining equation~(\ref{equ::inorm}) with Lemma~\ref{upperbound_marginaldistribution}, we have
		\begin{equation}
		\begin{aligned}
		\mathscr{I} &= \sum_i \hat{P_i}C_i - \frac{H(\mathbf{X})}{f} = \sum_i \hat{P_i}C_i - H(\mathbf{Y})\\
		& \geq  \sum_i \hat{P_i}C_i - H(\hat{P}) = D,
		\end{aligned}
		\end{equation}
		with equality if and only if the output process generated by $\phi$ approximates an i.i.d. process (Remark~\ref{remark::iid}).
		
		The minimum KL-divergence with average cost upper bounded by a specified average cost $C$, denoted by $D(C)$, was also studied in~\cite[Sec 5.1]{BochererThesis}. The pmf that achieves $D(C)$ is
		\begin{equation}
		\label{equ::de}
		\hat{P_i} = \frac{2^{-\mu C_i}}{\sum_j 2^{-\mu C_j}},\quad \sum_i \hat{P_i}C_i = C,
		\end{equation}
		when $C \leq \sum P_i C_i$, or $\mu \geq 1$. When $C > \sum P_i C_i$, by setting $\mu = 1$, we have $\hat{P_i} = P_i$, $ \sum_i \hat{P_i}C_i < C$, and 
		\begin{equation}
		D(C) = \sum_i \hat{P_i} C_i- H(\hat{P}) = 0.
		\end{equation}
		By comparing equations~(\ref{equ::iminfthm}) and~(\ref{equ::de}), we conclude that
		\begin{equation} 
		\mathscr{I}_\text{min}(f) = D(C),
		\end{equation}
		where
		\begin{equation}
		C = \sum_i \hat{P_i}C_i\quad f = \frac{H(\mathbf{X})}{-\sum_i \hat{P_i}\log_2 \hat{P_i}}\quad\hat{P_i} = \frac{2^{-\mu C_i}}{\sum_j 2^{-\mu C_j}},
		\end{equation}
		when $\mu \geq 1$, or equivalently when $C \leq \sum P_i C_i$ and  $f \geq f_\text{opt}$.
	\end{remark}
	
	The derivative $\mathrm{d}D(C)/\mathrm{d}C$ was  found in~\cite[Sec. 5.1]{BochererThesis}. Using the chain rule, we know that 
	\begin{equation}
	\frac{\mathrm{d}D(C)}{\mathrm{d}C} =\begin{cases} \frac{\mathrm{d}\mathscr{I}_\text{min}}{\mathrm{d}f}\frac{\mathrm{d}f}{\mathrm{d}C}\quad &\text{when $C \leq \sum P_i C_i$ ($\mu \geq 1$),}\\
	0 &\text{when $C > \sum P_i C_i$.}
	\end{cases}
	\end{equation}
	From (\ref{equ::averagecostandh}) we have
	\begin{equation}
	\frac{\mathrm{d}h}{\mathrm{d}C} =\frac{\mathrm{d}\frac{H(\mathbf{X})}{f}}{\mathrm{d}C}  = \mu \quad\Rightarrow\quad \frac{\mathrm{d}f}{\mathrm{d}C} = -\frac{\mu f^2}{H(\mathbf{X})}.
	\end{equation}
	By combining this with (\ref{equ:didmu1}), we have
	\begin{equation}
	\frac{\mathrm{d}D(C)}{\mathrm{d}C} =\begin{cases} 1-\mu \quad &\text{when $C \leq \sum P_i C_i$ ($\mu \geq 1$),}\\
	0 &\text{when $C > \sum P_i C_i$.}
	\end{cases}
	\end{equation}
	Therefore Proposition~\ref{prop::propertyiminf} allows us to recover the derivative of $D(C)$.\qed
	\begin{remark}
		In~\cite[Sec. V]{BochererAmjad},  a bound on the rate of a prefix-free variable-length DM code in the vicinity of $\mathscr{I}=0$ was given. Proposition~\ref{prop::propertyiminf} gives an explicit relationship between $\mathscr{I}$ and the code rate over a wider range of rates.\qed
	\end{remark}
	
	\begin{remark}
		In~\cite[Section 6.2.2]{Soriaga}, Soriaga considered the case where system requirements dictate that the expansion factor of the DM code encoder can not exceed $f_0$, where  $f_0 < f_{opt}$. In such a case, the code cannot be an optimal DM code for target distribution $\{P_i\}$. We may try to approximate an optimal DM code by designing a code with $f \leq f_0$ that minimizes the asymptotic normalized KL-divergence, $\lim_{l\rightarrow \infty} \frac{1}{l}D(Y_1^l || \tilde{Y}_1^l)$. We denote by $\mathscr{D}(f_0)$ the minimum possible value of this divergence. The relationship between $\mathscr{D}(f_0)$ and $f_0$ for a finite-order Markov target distribution was given in~\cite{Soriaga} and the result is also applicable to the i.i.d. case considered here. Since the asymptotic normalized  KL-divergence for a fixed $f$ is lower bounded by $\mathscr{I}_\text{min}(f)$ (Theorem~\ref{thm::nidtypeone}) and $\mathscr{I}_\text{min}(f)$ is strictly monotone decreasing when $f \leq f_0 < f_{opt}$ (Proposition~\ref{prop::propertyiminf}), we have 
		
		\begin{equation}
		\begin{aligned}
		\mathscr{D}(f_0)  &=\mathscr{I}_\text{min}(f_0) = \sum_i \hat{P_i}C_i - \frac{H(\mathbf{X})}{f_0} \\ 
		& = \sum_i \hat{P_i}\frac{(-\log_2 {\hat{P_i}}- \log_2 N)}{\mu}- \frac{H(\mathbf{X})}{f_0}\\
		&= \frac{-\sum_i \hat{P_i}\log_2 {\hat{P_i}}}{\mu}-\sum_i P_i\frac{\log_2 N}{\mu}-\frac{H(\mathbf{X})}{f_0}\\\\
		& = \frac{H(\mathbf{X})}{\mu f_0} - \frac{\log_2 N}{\mu} - \frac{H(\mathbf{X})}{f_0},
		\end{aligned}
		\end{equation}
		where $\mu$ and $N$ are constants such that $N = \sum_i 2^{-\mu C_i}$ and $\sum_i -\hat{P}_i \log \hat{P}_i=H(\mathbf{X})/f_0$, with $\hat{P}_i = \frac{1}{N} 2^{-\mu C_i}$.
		Because the code that achieves this lower bound with expansion factor $f_0$ is an optimal type-\Romannum{1} shaping code, based on the equivalence theorem we  can extend the result in~\cite[Section 6.2.2]{Soriaga} by concluding that this code is an optimal DM code for target distribution $\{\hat{P}_i\}$.
	\end{remark}

	\begin{remark}
		In Theorem~\ref{thm::nidtypeone}, we have shown that when $\mathscr{I} \rightarrow 0$, then $f\rightarrow f_\text{opt}$. This implies that the  GEF of the code satisfies
		\begin{equation}
		F = \frac{f \mathscr{I} + H(\mathbf{X})}{\log_2|\mathcal{Y}| } \rightarrow \frac{H(\mathbf{X})}{\log_2 |\mathcal{Y}|}.
		\end{equation}
		Similarly, as shown in Appendix~\ref{appen:type1}, when $F\rightarrow H(\mathbf{X})/\log_2 |\mathcal{Y}|$ (or, equivalently, when total cost $T\rightarrow H(\mathbf{X})$), then $f\rightarrow f_\text{opt}$ and 
		\begin{equation}
		\mathscr{I}= \frac{F - \frac{H(\mathbf{X})}{\log_2 |\mathcal{Y}|}}{f}\rightarrow 0.
		\end{equation}
		In view of the equivalence between asymptotic normalized KL-divergence and $\mathscr{I}$, these observations extend Theorem~\ref{thm::geflowerbound} by providing  bounds on asymptotic normalized KL-divergence in the vicinity of $F = H(\mathbf{X})/ \log_2 |\mathcal{Y}|$.\qed
	\end{remark}
	%%%%%%%%%%%%%%%%%%%%%%%%%%%%%%%%%%%%%%%%%%%%%%%%%
	%%%%%%%%%%%%%%%%%%%%%%%%%%%%%%%%%%%%%%%%%%%%%%%%%
	\section{Experimental Results}
	\label{sec:experiment}
	%%%%%%%%%%%%%%%%%%%%%%%%%%%%%%%%%%%%%%%%%%%%%%%%%
	%%%%%%%%%%%%%%%%%%%%%%%%%%%%%%%%%%%%%%%%%%%%%%%%%
	\subsection{Optimal Data Shaping Code for MLC Flash Memory}
	\label{sec:nand}
\begin{figure}[h]
		\centering
		\includegraphics[width=1\columnwidth]{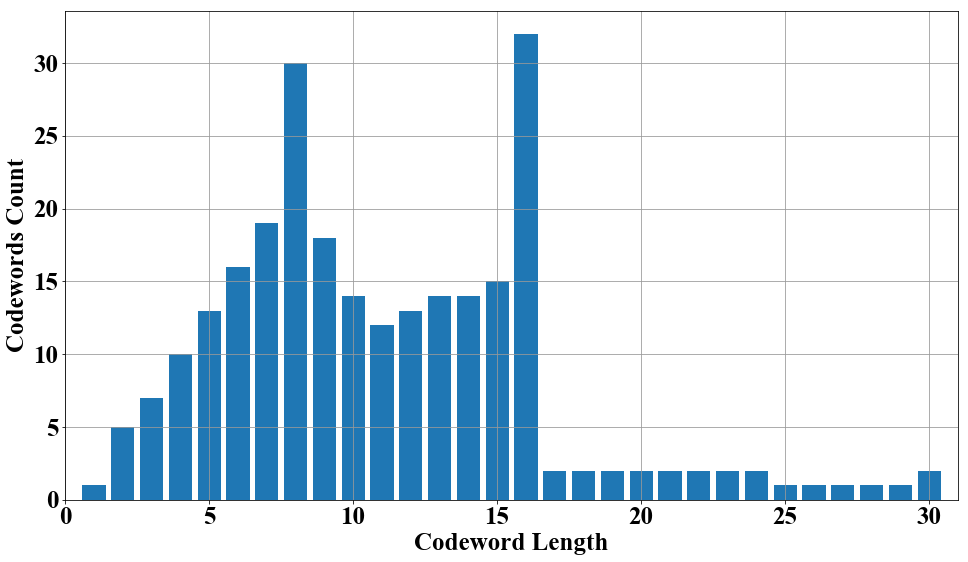}
		\caption{Codeword length distribution of Varn code with the codebook size $K=256$ for English-language text. }
		\label{fig::length_count}
	\end{figure}

	We evaluated the performance of  shaping codes on a multilevel-cell (MLC) NAND flash memory. 
	%NAND flash memory is a non-volatile storage medium with memory elements, or cells, which are electrically programmable and erasable. 
	In MLC flash, the cells are arranged in a rectangular array (also called a \textit{block}) and each row of cells is called a \textit{wordline}. The cells can be programmed to four different voltage levels, denoted $\{0,1,2,3\}$, so each cell can store two bits of information.  It was shown in \cite{LiuSieGC16}, \cite{LiuNVMW2016} that MLC flash memory can be modeled as a costly channel with alphabet $\{0,1,2,3\}$, where  the cost of the erase level 0 can be taken to be  $C_0=0$. Using the methodology described in \cite{LiuSieGC16}, the cost vector for the memory was found empirically to be 
	\begin{equation}
	\mathcal{C} = [0, 0.58, 0.87, 1.29].
	\end{equation}
	From Theorem~\ref{opt_expansion}, we know that the total   cost is a decreasing function of the expansion factor. To assess the performance of optimal shaping, and to permit a comparison to the direct-shaping code in \cite{LiuSieGC16}, \cite{LiuNVMW2016},
	%The cost vector for this channel is $\mathcal{C} = \{0,0.59,1.07,1.43\}$ (for details of this model, see \cite{LiuSieGC16}). 
	we applied a rate-1, type-\Romannum{1} shaping code to the ASCII representation of the English-language text of \textit{The Count of Monte Cristo}. The ``optimal'' shaping scheme was designed according to the principles suggested by the equivalence theorem and separation theorem. We first compressed the file using the LZ77 algorithm. The observed compression rate was $g = 1/2.740$. We then used Theorem~\ref{performance:opt} to compute the target symbol occurrence probabilities of  a shaping  code that minimizes average cost for a uniform i.i.d. source, a cost vector $\mathcal{C}$, and the expansion factor $f'=f/g= 2.740$.  The resulting symbol occurrence probability distribution was given by 
	\begin{equation}
	\hat{P}=[ 0.8606,  0.0989, 0.0335,  0.0070].
	\end{equation} 
	Using Theorem~\ref{thm:typeonetypetwo}, we computed the costs for the equivalent code that minimizes total cost, yielding the cost vector 
	\begin{equation}
	\mathcal{C}'=  [0.2167, 3.3378, 4.8983,  7.1585].   
	\end{equation}
	We constructed a Varn code with codebook size $K=256$ based on the cost vector $\mathcal{C}'$. This code is a length-8, type-\Romannum{2}  shaping code and the concatenation of the compression and the Varn code is a rate-1, type-\Romannum{1} shaping code. The expansion factor of the Varn code is 2.768, which is close to the expansion factor of the optimal type-\Romannum{2} shaping code for cost vector $\mathcal{C}'$, where $f_{opt} = 2.737$. Its codeword length distribution is shown in Fig.~\ref{fig::length_count}.
	
\begin{figure}[h]
		\centering     %%% not \center
		\subfigure[]{\label{eng:fig1}\includegraphics[width=0.49\columnwidth]{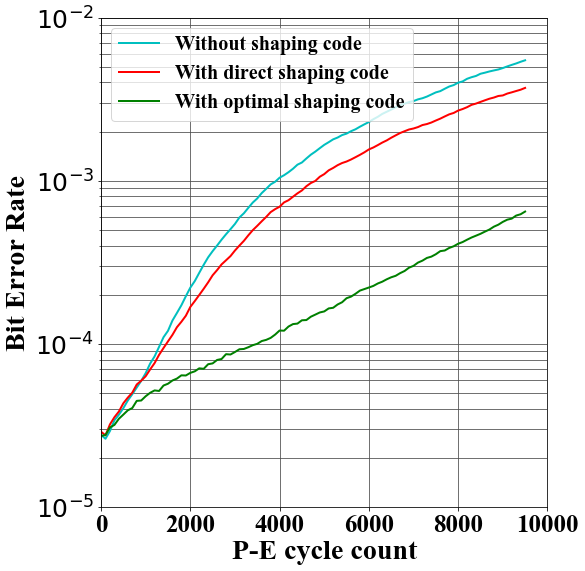}}
		\subfigure[]{\label{eng:fig2}\includegraphics[width=0.49\columnwidth]{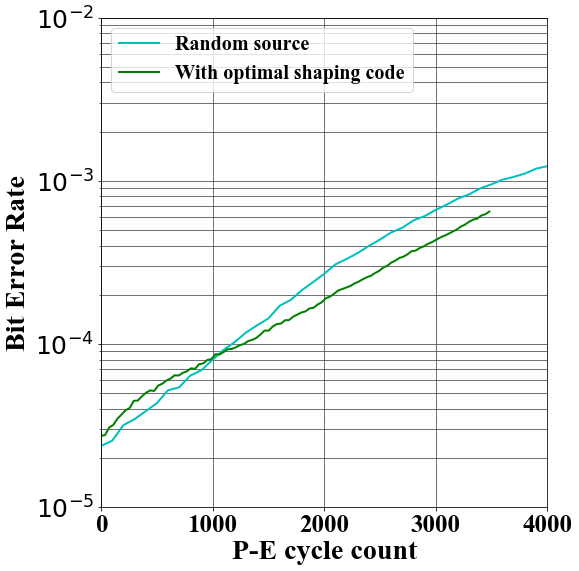}}
		\vspace{-0.75em}
		\caption{BER performance for English-language text.}
	\end{figure}

	To characterize the performance of the designed shaping code, we performed a program/erase (P/E) cycling experiment on the MLC flash memory by repeating the following steps, which collectively represent one P/E cycle. The experiment was conducted with the uncoded source data, and then with the output data from the shaping code.
	\begin{itemize}
		\item Erase the MLC flash memory block.
		\item Program the MLC flash memory.
		\item For each successive programming cycle, ``rotate" the data, so the data that was written on the $i$th wordline is  written on the $(i+1)$st wordline, wrapping around the last wordline to the first wordline.
		\item After every 100 P/E cycles, erase the block and program pseudo-random data. Then perform a read operation, record bit errors, and calculate the bit error rate.
	\end{itemize}
	Fig.~\ref{eng:fig1} shows the average bit error rates (BERs) for the uncoded source data, the direct shaping code \cite{LiuSieGC16}, and the optimal shaping code. The results indicate that the optimal shaping code  provides a significant increase in the memory lifetime compared to no shaping and direct shaping.

	As a way of comparing the performance of optimal shaping to that of data compression alone, we rescaled the P/E cycle count of the shaping code by the compression ratio 2.740 and compared the result to P/E cycling of pseudo-random data. This corresponds to a BER comparison based upon the total amount of source data stored in the memory. The results, shown in Fig.~\ref{eng:fig2}, indicate that the performance of optimal shaping is superior to data compression alone as a function of total source data written. 
	
\begin{figure}[h]
		\centering
		\includegraphics[width=1\columnwidth]{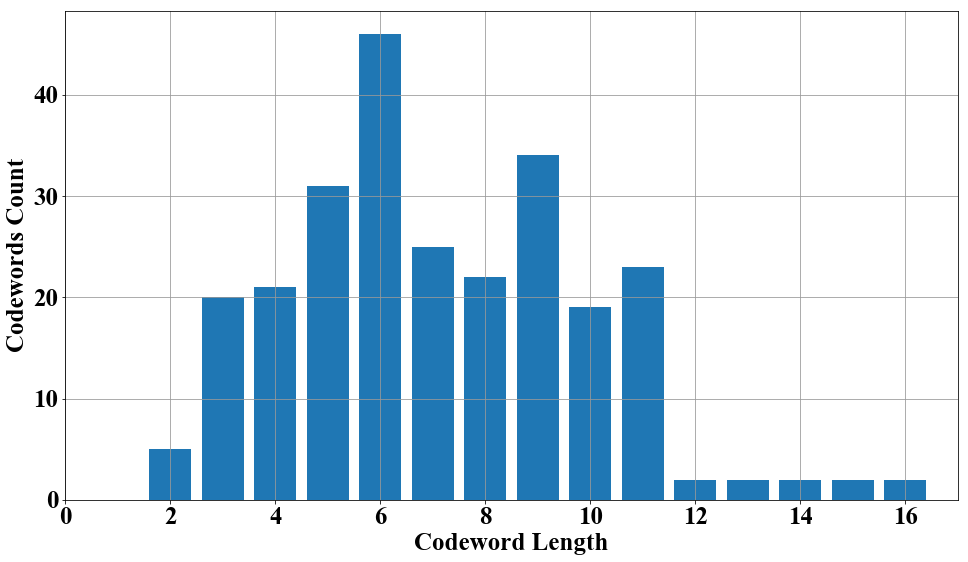}
		\caption{Codeword length distribution of Varn code with codebook size $K=256$ for Chinese-language text.}
		\label{fig::length_luxun}
	\end{figure}

\begin{figure}[h]
		\centering     %%% not \center
		\subfigure[]{\label{chn:fig1}\includegraphics[width=0.49\columnwidth]{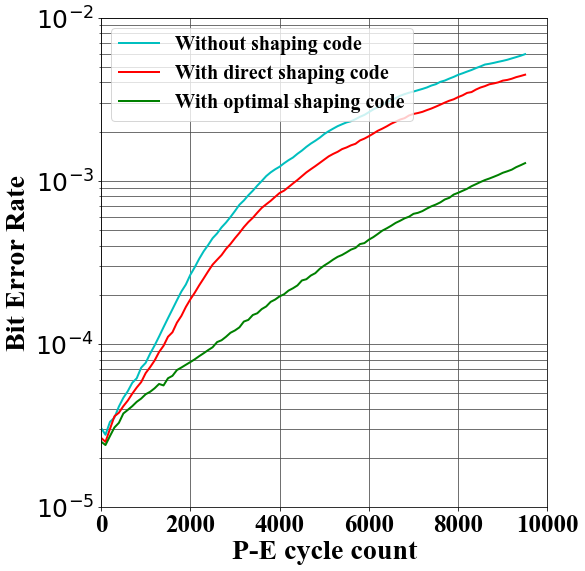}}
		\subfigure[]{\label{chn:fig2}\includegraphics[width=0.49\columnwidth]{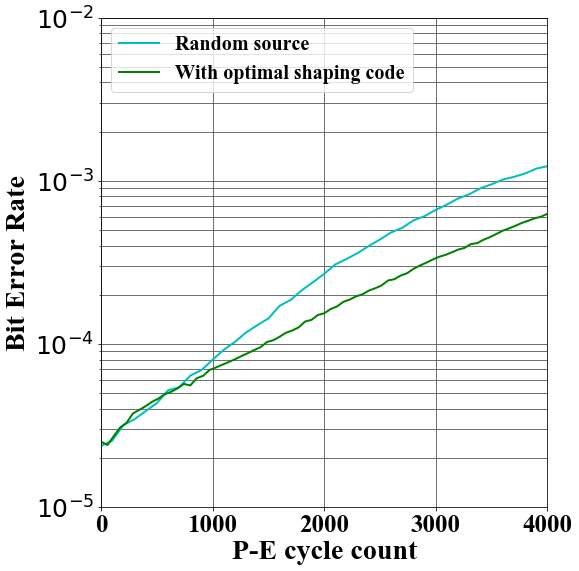}}
		\vspace{-0.75em}
		\caption{BER performance for Chinese-language text.}
	\end{figure}

	A similar experiment was conducted for a Chinese-language text, \textit{Collected Works of Lu Xun, Volumes 1--4}, represented using UTF-16LE encoding. We constructed a Varn code with codebook size $K=256$ based on the cost vector
	\begin{equation}
	\mathcal{C}'=  [0.4222, 2.6647,3.7860, 5.4099].   
	\end{equation}
	The expansion factor of the the Varn code was 1.751, which is close to the expansion factor of the optimal type-\Romannum{2} shaping code,  $f_{opt} = 1.759$. Its codeword length distribution is shown in Fig.~\ref{fig::length_luxun}. The BER results are shown in  Fig.~\ref{chn:fig1} and Fig.~\ref{chn:fig2}.

	
	%%%%%%%%%%%%%%%%%%%%%%%%%%%%%%%%%%%%%%%%%%%%%%%%%
	%%%%%%%%%%%%%%%%%%%%%%%%%%%%%%%%%%%%%%%%%%%%%%%%%
	
	\subsection{Varn Codes for Distribution Matching}
	Remark~\ref{remark:varnasdm} shows that the upper bound on the generalized expansion factor of Varn codes decreases as the codebook size increases.  This  suggests that as the codebook size of a Varn code increases,  the  approximation to an optimal DM code should improve. In this subsection, we empirically tested this premise by constructing Varn codes with codebook size $K = 100$, 1000 and 10000, respectively, for a target distribution $\{P_0, P_1\} = \{\frac{2}{3}, \frac{1}{3}\}$. The measure of goodness we used here was similar to the serial test in~\cite[Section 2.11]{NIST}, namely  KL-divergences for patterns of increasing length. Codeword sequences with 10000 codewords were generated using the random number sequence collected from \cite{random}. The first 71514 bits in codeword sequences were used for comparison (71514 was the length of the codeword sequence generated by the Varn code with codebook size $K = 100$). The probability of occurrence of length 1, 2 and 3 patterns was calculated. For example, we define the probability of occurrence of `10' ($P_{10}'$) and `101' ($P_{101}'$) in codeword sequence $y_1^l$ as
	
	\begin{equation}
	P_{10}' = \frac{\{\text{the number of subsequences $y_i^{i+1}= $`10'$$\}}}{l-1},
	\end{equation}
	\begin{equation}
	P_{101}' = \frac{\{\text{the number of subsequences $y_i^{i+2} = $`101'$$\}}}{l-2}.
	\end{equation}
	The first-, second-, and third-order  KL-divergences between $P'$ and distribution  $\{P_0, P_1\} = \{\frac{2}{3}, \frac{1}{3}\}$ were calculated, using the following definitions:
	\begin{equation}
	I_1 = \sum_{i \in \{0,1\}} P'_{i} \log_2 \frac{P'_{i}}{P_i}
	\end{equation}
	\begin{equation}
	I_2 = \sum_{i \in \{0,1\}} \sum_{j \in \{0,1\}} P'_{ij} \log_2 \frac{P'_{ij}}{P_i P_j}
	\end{equation}
	\begin{equation}
	I_3 = \sum_{i \in \{0,1\}} \sum_{j \in \{0,1\}} \sum_{k\in \{0,1\}} P'_{ijk} \log_2 \frac{P'_{ijk}}{P_i P_jP_k}.
	\end{equation} 
	The results are shown in Table~\Romannum{1}. The divergences decrease as $K$ increases, indicating that the approximation  to an i.i.d. sequence with target distribution $\{P_0, P_1\} = \{\frac{2}{3}, \frac{1}{3}\}$ is improving.
	
	\begin{table}[h]
		\centering
		\label{table::Idivergence}
		\begin{tabular}{cccccc}
			\hline
			& $P_0'$ & $I_1$ & $I_2$ & $I_3$ \\ \hline
			$K=100$          & 0.6447              &0.0015  & 0.0032 & 0.0055     \\
			$K=1000$     & 0.6498                    &  0.00091   & 0.0018 & 0.0027     \\
			$K=10000$    &0.6602                    &  0.00014  & 0.00027    & 0.00028 \\ \hline
		\end{tabular}
		\caption{First- , second-,  and third-order  KL-divergence}
	\end{table}
	
	%%%%%%%%%%%%%%%%%%%%%%%%%%%%%%%%%%%%%%%%%%%%%%%%%
	%%%%%%%%%%%%%%%%%%%%%%%%%%%%%%%%%%%%%%%%%%%%%%%%%
	\section{Conclusion}
	\label{sec:conclude}
	In this paper, we studied information-theoretic properties and performance limits of a general class of shaping codes. We determined the asymptotic symbol occurrence probability distribution, and used it to determine the minimum achievable average   cost for a type-\Romannum{1} shaping code.  Using these results, we determined the minimum total   cost and  optimal expansion factor for a type-\Romannum{2} shaping code.  A consequence of this analysis is an equivalence theorem, stating that a type-\Romannum{1} shaping code with a given expansion factor and a cost vector can be realized by a type-\Romannum{2} shaping code. We then proved a separation theorem stating that optimal shaping can be achieved  by a concatenation of optimal lossless compression and optimal shaping for a uniform i.i.d. source. Experimental results showed that optimal shaping can provide a significant increase in flash memory lifetime when  applied to English-language  and Chinese-language texts, providing total data capacity greater than that achieved by data compression alone.
	
	We also studied properties of prefix-free variable-length distribution matching (DM) codes from the perspective of shaping. We characterized optimal DM codes in terms of the asymptotic normalized divergence and showed that when the divergence equals zero, a DM code encoder generates a codeword sequence that looks i.i.d., with symbol occurrence probability equal to the target distribution. 
	We showed that optimal type-\Romannum{2}  shaping codes can be used to construct optimal DM codes. This suggested the definition of the \text{generalized expansion factor} as a performance measure for DM codes and implied a separation theorem for DM codes. We also established the relationship between the generalized expansion factor and the informational divergence of a DM code. The relationship between the type-\Romannum{1} shaping problem and the minimization of normalized informational divergence was also studied. Simulation results showed an increase in distribution matching performance of Varn codes designed for a Bernoulli distribution as the codebook size increases.

	%%%%%%%%%%%%%%%%%%%%%%%%%%%%%%%%%%%%%%%%%%%%%%%%%
	%%%%%%%%%%%%%%%%%%%%%%%%%%%%%%%%%%%%%%%%%%%%%%%%%
	
	\section*{Acknowledgment}
	This work was supported in part by National Science Foundation (NSF) Grant CCF-1619053.
	
	%%%%%%%%%%%%%%%%%%%%%%%%%%%%%%%%%%%%%%%%%%%%%%%%%
	%%%%%%%%%%%%%%%%%%%%%%%%%%%%%%%%%%%%%%%%%%%%%%%%%

	\begin{appendices}
		\section{PROOF OF LEMMA~\ref{lemma:10}}
		\label{appen:1}
		\begin{IEEEproof}
			In this proof, without loss of generality, we will assume $q=1$.
			First, we evaluate the expectation of the sequence of random variable $\{N_i(\phi(X_1^{M_l}))\}_{l=1}^{\infty}$. 
			Combining Lemma~\ref{lemma::wald} with equation (\ref{equ:walds}), we have
			\begin{equation}
			\begin{split}
			\lim_{l\rightarrow \infty} \frac{1}{l} E(N_i(\phi(X_1^{M_l}))) &= \lim_{l\rightarrow \infty}\frac{1}{l}E(N_i(\phi(X)))E(M_l)\\
			&=E(N_i(\phi(X)))\lim_{l \rightarrow \infty}\frac{1}{l}E(M_l)\\ &= E(N_i(\phi(X)))\frac{1}{E(L)}.
			\end{split}
			\end{equation}
			Similarly, we have
			\begin{equation}
			\begin{split}
			\lim_{l\rightarrow \infty} &\frac{1}{l} E(N_i(\phi(X_1^{M_l-1}))) \\& = \lim_{l\rightarrow \infty} \frac{1}{l} E(N_i(\phi(X_1^{M_l}))-N_i(\phi(X_{M_l}))) \\&= E(N_i(\phi(X)))\frac{1}{E(L)} -\lim_{l\rightarrow \infty}\frac{1}{l}E(N_i(\phi(X_{M_l})))\\
			& = E(N_i(\phi(X)))\frac{1}{E(L)},
			\end{split}\end{equation}
			where $\lim_{l\rightarrow \infty}\frac{1}{l}E(N_i(\phi(X_{M_l}))) = 0$ follows from Lemma~\ref{lemma:11}.
			
			By definition,
			\begin{equation}
			E(N_i(\phi(X_1^{M_l})))=\sum_{y_1^l}\sum_{x_1^{M_l}\in\mathcal{G}_{\phi}(y_1^l)}P(x_1^{M_l})N_i(\phi(x_1^{M_l})).
			\end{equation}
			Since $S_{M_l}>l$, we have $N_i(\phi(x_1^{M_l}))\geq N_i(y_1^l)$ and $E(N_i(y_1^l))$ can be bounded as follows
			\begin{equation}\label{Asym_upperbound}
			\begin{split}
			E(N_i(Y_1^l)) & = \sum_{y_1^l}N_i(y_1^l) Q(y_1^l)\\ 
			& = \sum_{y_1^l} N_i(y_1^l)\sum_{x_1^{M_l}\in\mathcal{G}_{\phi}(y_1^l)}P(x_1^{M_l}) \\
			& \leq  \sum_{y_1^l} N_i(\phi(x_1^{M_l}))\sum_{x_1^{M_l}\in\mathcal{G}_{\phi}(y_1^l)}P(x_1^{M_l}) \\
			& =\sum_{y_1^l}\sum_{x_1^{M_l}\in\mathcal{G}_{\phi}(y_1^l)}P(x_1^{M_l})N_i(\phi(x_1^{M_l})) \\
			& = E(N_i(\phi(X_1^{M_l}))).
			\end{split}
			\end{equation}
			Similarly, $N_i(\phi(x_1^{M_l-1}))\leq N_i(\phi(y_1^l))$ and $E(N_i(Y_1^l))$ is lower bounded by
			\begin{equation}\label{Asym_lowerbound}
			\begin{split}
			E(N_i(Y_1^l)) &=  \sum_{y_1^l} N_i(y_1^l)\sum_{x_1^{M_l}\in\mathcal{G}_{\phi}(y_1^l)}P(x_1^{M_l})\\
			& \geq \sum_{y_1^l}\sum_{x_1^{M_l}\in\mathcal{G}_{\phi}(y_1^l)}P(x_1^{M_l})N_i(\phi(x_1^{M_l-1}))\\
			& = E(N_i(\phi(X_1^{M_l-1}))).
			\end{split}
			\end{equation}
			Equations (\ref{Asym_upperbound})  and (\ref{Asym_lowerbound}) imply that
			\begin{equation}
			\begin{split}
			\limsup_{l\rightarrow \infty}\frac{1}{l}E(N_i(Y_1^l)) &\leq \liminf_{l\rightarrow \infty}\frac{1}{l}E(N_i(\phi(X_1^{M_l})))\\
			&= E(N_i(\phi(X)))\frac{1}{E(L)}
			\end{split}
			\end{equation}
			and
			\begin{equation}
			\begin{split}
			\liminf_{l\rightarrow \infty}\frac{1}{l}E(N_i(Y_1^l)) & \geq \limsup_{l\rightarrow \infty}\frac{1}{l}E(N_i(\phi(X_1^{M_l-1})))\\ & = E(N_i(\phi(X)))\frac{1}{E(L)}.
			\end{split}
			\end{equation}
			Thus we conclude  that
			\begin{equation}
			\hat{P_i}=\lim_{l\rightarrow \infty}\frac{1}{l}E(N_i(Y_1^l)) = E(N_i(\phi(X)))\frac{1}{E(L)}.
			\end{equation}
		\end{IEEEproof}
		
		\section{PROOF OF THEOREM \ref{opt_expansion}} 
		\label{appen:2}
		\begin{IEEEproof}
			We solve the optimization problem:
			\begin{equation}
			\begin{aligned}
			& \underset{\hat{P}_i,f}{\text{minimize}}
			& & f\sum_{i}\hat{P}_iC_i\\
			& \text{subject to}
			& & H(\hat{Y})\geq\frac{H(\mathbf{X})}{f}\\
			&  && \sum_{i}\hat{P}_i=1.
			\end{aligned}
			\end{equation} 	
			We divide this optimization problem into two parts. First we fix expansion factor $f$ and find the minimum achievable average cost using Theorem~\ref{performance:opt}. Then we find the optimal $f$ to minimize the total   cost. The optimization problem then becomes:
			
			\begin{equation}\label{equ:optimizetotal}
			\begin{aligned}
			& \underset{f}{\text{minimize}}\quad \underset{\hat{P}_i}{\text{minimize}}
			& & \sum_{i}f\hat{P}_iC_i\\
			& \text{subject to}
			& & H(\hat{Y})\geq \frac{H(\mathbf{X})}{f}\\
			&  && \sum_{i}\hat{P}_i=1.
			\end{aligned}
			\vspace{-0.8em}
			\end{equation}
			For fixed $f$, the symbol occurrence probability corresponding to the minimum average cost is
			\begin{equation}
			\label{result:p}
			P_i'=\frac{1}{N}2^{-\mu C_i},
			\end{equation}
			where $\mu$ is a constant such that 
			\begin{equation}
			\label{constrain:f}
			\frac{1}{f}=\frac{-\sum_{i}P_i'\log_2P_i'}{\log_2|\mathcal{X}|} 
			\end{equation}
			and $N$ is the normalization factor
			\begin{equation}
			N= \sum_{i}2^{-\mu C_i}.
			\end{equation}
			
			Treating $f$ as a function of $\mu$,  and calculating $\mathrm{d}f/\mathrm{d}\mu$, we see that $f(\mu)$ is a monotone increasing function of $\mu$, with $f(0)=\frac{H(\mathbf{X})}{\log_2 |\mathcal{Y}|}$.  Specifically, 
			replacing the $P_i'$ in (\ref{constrain:f}) by (\ref{result:p}) gives
			\begin{equation}
			\label{constrain:two}
			f=\frac{NH(\mathbf{X}) }{\left(\sum_i\mu C_i 2^{-\mu C_i}\right)+N\log_2 N}.
			\end{equation}
			The derivative of $f$ with respect to $\mu$ is 
			\begin{equation}
			\label{equ:dfdmu}
			\frac{\mathrm{d}f}{\mathrm{d}\mu} =\frac{\mu\ln 2 H(\mathbf{X}) \sum_{i< j}2^{-\mu (C_i+C_j)} (C_i - C_j )^2}{[(\sum_i\mu C_i 2^{-\mu C_i})+N\log_2 N]^2},
			\end{equation}	
			which is easily seen to be positive for $\mu >0$.
			
			The optimization problem is then equivalent to:
			\begin{equation}\label{equ:optimize3}
			\begin{aligned}
			&\text{minimize}
			& & T(\mu)=H(\mathbf{X})\frac{\sum_i C_i 2^{-\mu C_i}}{\sum_i \mu C_i 2^{-\mu C_i}+ N \log_2 N}\\
			& \text{subject to}
			& & \mu\geq 0. \\
			\end{aligned}
			\end{equation}
			
			Calculating $\mathrm{d}T/\mathrm{d}\mu$, we see  that its sign is the negative of the sign of $\log_2 N$. 
			Specifically, we find that 
			\begin{small}
				\begin{equation}
				\frac{\mathrm{d}T}{\mathrm{d}\mu} =-\ln 2 H(\mathbf{X})\log_2N \frac{N\sum_iC_i^2 2^{-\mu C_i}-(\sum_i C_i2^{-\mu C_i})^2}{(\sum_i \mu C_i 2^{-\mu C_i}+N\log_2 N)^2}.
				\end{equation}
			\end{small}
			Observe that 
			\begin{equation}
			\begin{aligned}
			&N\sum_i C_i^2 2^{-\mu C_i}-(\sum_i C_i2^{-\mu C_i})^2 \\ &=  \sum_{i,j}C_i^22^{-\mu(C_i+C_j)} - \sum_iC_i^2 2^{-2 C_i}-\sum_{i<j}2C_iC_j 2^{-\mu(C_i+C_j)}\\
			&=\sum_{i<j}(C_i^2+C_j^2-2C_iC_j)2^{-\mu(C_i+C_j)} \\
			&= \sum_{i<j}(C_i-C_j)^22^{-\mu(C_i+C_j)} >0.
			\end{aligned}
			\end{equation}
			Therefore, 
			\begin{equation}
			\label{equ:dtdmu}
			\frac{\mathrm{d}T}{\mathrm{d}\mu} = - \ln 2H(\mathbf{X})\log_2N\frac{\sum_{i<j} (C_i-C_j)^2 2^{-\mu(C_i+C_j)}}{(\sum_i \mu C_i 2^{-\mu C_i}+ N \log_2 N)^2}.
			\end{equation}
			The claimed relationship between the sign of $\mathrm{d}T/\mathrm{d}\mu$ and the sign of $\log_2 N$ is then evident.

			It follows that if $C_1 = 0$, then $N=\sum_i 2^{-\mu C_i}>1$, implying  that $T(\mu)$ is a monotone decreasing function on $[0,\infty)$.
			On the other hand, if $C_1 > 0$, then when $\mu=0$,  we have $N=|\mathcal{Y}|$, so $\log_2{N} > 0$.
			Since $\mathrm{d}N/\mathrm{d}\mu <0$ for all $\mu$, we conclude that $T$ will decrease as $\mu$ increases, reaching a minimum at $N=1$. Beyond that point,   $\mathrm{d}N/\mathrm{d}\mu > 0$.
			
			%Since
			%			\begin{equation}
			%			\frac{\mathrm{d}N}{\mathrm{d}\mu} = -\sum_i C_i 2^{-\mu C_i} < 0 
			%			\end{equation}
			%			This impies $\frac{\mathrm{d}F}{\mathrm{d}\mu} < 0$ when $\mu$ is small, meaning $F$ will decrease as $\mu$ increase, or $f$ increase. It will reach its minimum when $N = 1$. After that $\frac{\mathrm{d}F}{\mathrm{d}\mu} > 0$.
			Thus, the corresponding expansion factor that achieves the minimum total cost is 
			\begin{equation}
			f= \frac{H(\mathbf{X})}{-\sum_i \hat{P}_i \log_2 \hat{P}_i}
			\end{equation}
			where $\hat{P}_i=2^{-\mu C_i}$,  and $\mu$ is a positive constant satisfying $\sum_i 2^{-\mu C_i}=1$. 
		\end{IEEEproof}
		
		%\section{ANOTHER PROOF OF THEOREM~\ref{thm:typeonetypetwo}}
		%	\label{appen:typeonetypetwo}
		
		\section{PROOF OF THEOREM \ref{separation}}
		\label{appen:3}
		Before proving the separation theorem, we first design a type-\Romannum{2}  shaping code for the uniform source. The \textit{average codeword cost} $C(K)$ for a Varn code with codebook size $K$ is bounded by
		\begin{equation}
		\log_2 K / \mu \leq C(K) \leq \log_2 K / \mu + \max_i C_i 
		\end{equation}
		where $\mu$ is a constant such that 
		\begin{equation}
		\sum_i 2^{-\mu C_i}  = 1.
		\end{equation}
		Even though the average codeword cost is bounded, it is not guaranteed that the cost of every codeword satisfies the same bound. In order to ensure this,  we instead use a modified design which we refer to as a modified Varn code.
		Given the binary alphabets $\{0,1\}$ and $\mathcal{Y}$, a tree-based variable-length modified Varn code $\phi: \mathcal{X}^K \rightarrow \mathcal{Y}^*$ is designed as follows:
		\begin{itemize}
			\item Let $\nu = [(2^K - 1) \mod (|\mathcal{Y}|-1)]$.
			\item If $\nu > 0$, let $\delta = |\mathcal{Y}| - 1 -\nu$. Else if $\nu = 0$, let $\delta =0$. Set $M = 2^K + \delta$.
			\item Design an exhaustive Varn code with codebook size $M$.
			\item Trim down the tree by getting rid of the $\delta$ branches with largest cost.
		\end{itemize}
		We note  that $\delta \leq |\mathcal{Y}| - 2$.  
		
		The following lemma gives an upper bound on the codeword cost in a modified Varn code.
		\begin{lemma}
			Every codeword of the modified Varn code has cost upper bounded by
			\begin{equation}
			\label{reverseTunstallBound}
			W_i \leq \log_2 M / \mu + \max_i C_i.
			\end{equation}\qed
			%\vspace{-3em}
		\end{lemma}
		\begin{IEEEproof}
			Consider the internal node that was the last one to be expanded and  suppose it has cost $W_0$. The cost of any leaf node is larger than $W_0$, so
			\begin{equation}
			\label{RTC:lower}
			W_i \geq W_0.
			\end{equation}
			Since this internal node is the last one to be expanded, its cost is larger than that of any other internal node, implying
			\begin{equation}
			\label{RTC:upper}
			W_i \leq W_0 + \max C_i.
			\end{equation}
			Since the tree is full, it is easy to check that 
			\begin{equation}
			\label{RTC:tree}
			\sum_{i} 2^ {-\mu W_i} = 1.
			\end{equation}
			Combining equations (\ref{RTC:lower}) and (\ref{RTC:tree}), we have 
			\begin{equation}
			1 = \sum_{i} 2^{-\mu W_i} \leq M 2^{-\mu W_0}.
			\end{equation}
			This implies
			\begin{equation}
			W_0 \leq \log_2 M  / \mu
			\end{equation}
			and
			\begin{equation}
			W_i \leq W_0 + \max C_i \leq \log_2 M / \mu + \max C_i.
			\end{equation}
		\end{IEEEproof}
		
		Now we are ready to prove the separation theorem.
		\begin{IEEEproof}
			Let's first define a constant
			\begin{equation}
			D = \ceilenv{\frac{\log—_2 |\mathcal{X}|+2}{H(\mathbf{X})}}.
			\end{equation}
			For any given $\gamma > 0$, define
			\begin{equation}
			\epsilon = \min\{\frac{\mu \gamma}{2H(\mathbf{X})D},\frac{1}{D} , \frac{\mu\gamma}{16}\}
			\end{equation}
			and $\epsilon' = \frac{\mu\gamma}{16}$. Consider the typical set $A_\epsilon^{(q)}$ with respect to an i.i.d. source with entropy $H(\mathbf{X})$. There exists positive integer $Q_1$ such that when $q> Q_1$, $\text{Pr}\{A_\epsilon^{(q)}\} > 1 - \epsilon$. There are $\leq 2^{q(H(\mathbf{X})+\epsilon)}$ length-$q$ sequences $x^q$ in $A_\epsilon^{(q)}$, so we can use no more than $\ceilenv{q(H+\epsilon)+1}$ bits to index them.  We prefix all these sequences by a 0, giving a total length of $\ceilenv{ q(H(\mathbf{X})+\epsilon)+2}$ to represent each sequence in $A_\epsilon^{(q)}$. Similarly, we can index each sequence not in $A_\epsilon ^{(q)}$ by using $\ceilenv{q\log|\mathcal{X}|}$ bits. Prefixing these indices by 1, we have a prefix-free code $\psi_1$ for all sequences in $\mathcal{X}^q$.
			
			Now we construct a length-$\ceilenv{q(H(\mathbf{X})+ \epsilon)+2}$ modified Varn code $\psi_2: \mathcal{X}^{\ceilenv{q(H(\mathbf{X})+ \epsilon)+2}} \rightarrow \mathcal{Y}^*$. We use this code to encode the codeword sequence generated by $\psi_1$. For every codeword in $A_\epsilon^{(q)}$, the cost is upper bounded by $\log_2 M /\mu + \max_i C_i$. For every codeword in the complement of $A_\epsilon^{(q)}$, which we denote by $B_\epsilon^{(q)}$, $\ceilenv{\frac{\ceilenv{q\log|\mathcal{X}|+1}}{\ceilenv{ q(H(\mathbf{X})+\epsilon)+2}}}$ codewords in $\psi_2$ are needed.  The total cost is upper bounded by $(\ceilenv{\frac{\ceilenv{q\log|\mathcal{X}|+1}}{\ceilenv{ q(H(\mathbf{X})+\epsilon)+2}}}) (\log_2 M /\mu + \max_i C_i)$, where  
			\begin{equation}
			\ceilenv{\frac{\ceilenv{q\log|\mathcal{X}|+1}}{\ceilenv{ q(H(\mathbf{X})+\epsilon)+2}}} \leq \ceilenv{\frac{q\log|\mathcal{X}|+2q}{ qH(\mathbf{X})}} = D.
			\end{equation}
			Consider the concatenation of $\psi_1$ and $\psi_2$, denoted $\Psi = \Psi_2 \circ \Psi_1$, with $\Psi (\mathcal{X}^q) = \psi_2(\psi_1(\mathcal{X}^q))$. The total cost is 
			%$\Psi (\mathcal{X}^q) = \psi_2(\psi_1(\mathcal{X}^q))$, the total   cost is 
			\begin{equation}
			\begin{split}
			T&(\Psi ) = \frac{1}{q}(\sum_{x^q}P(x^q)W(x^q))\\ 
			& = \frac{1}{q}(\sum_{x^q \in A_\epsilon^{(q)}}P(x^q)W(x^q) + \sum_{x^q \in A_\epsilon^{(q)C}}P(x^q)W(x^q))\\
			& \leq \frac{1}{q}[\sum_{x^q \in A_\epsilon^{(q)}}P(x^q)(\log_2 M /\mu + \max_i C_i) \\ &+ \sum_{x^q \in A_\epsilon^{(q)C}}P(x^q)D (\log_2 M /\mu + \max_i C_i)]\\
			& =\frac{1}{q}[\text{Pr}\{A_\epsilon^{(q)}\}(\log_2 M /\mu + \max_i C_i) \\&+ \text{Pr}\{A_\epsilon^{(q)C}\}D (\log_2 M /\mu + \max_i C_i)]\\
			& < \frac{1}{q}[(\log_2 M /\mu + \max_i C_i) \\&+ \epsilon D (\log_2 M /\mu + \max_i C_i)]\\
			& = (\frac{\log_2 M}{q\mu} + \frac{\max_i C_i}{q})(1+\epsilon D)
			\end{split}
			\end{equation}
			where
			\begin{equation}
			M = 2^{\ceilenv{q(H(\mathbf{X})+ \epsilon)+2}}+\delta.
			\end{equation}
			Since $\delta \leq |\mathcal{Y}| - 2$, we can bound $M$ by
			\begin{equation}
			M \leq 2^{q(H(\mathbf{X})+ \epsilon)+3}+|\mathcal{Y}|
			\end{equation}
			and by L'H\"{o}pital's rule, we have
			\begin{equation}
			\lim_{q\rightarrow\infty} \frac{\log_2 ( 2^{q(H(\mathbf{X})+ \epsilon)+3}+|\mathcal{Y}|)}{q}= H(\mathbf{X})+ \epsilon.
			\end{equation}
			Therefore, there exists positive integer $Q_2$ such that, when $q> Q_2$,
			\begin{equation}
			\frac{\log_2 M}{q} < H(\mathbf{X}) + \epsilon + \epsilon'.
			\end{equation}
			Thus, when $q> \max\{Q_1,Q_2\}$, the total   cost of $\Psi$ is upper bounded by
			
			\begin{equation}
			T(\Psi ) <(\frac{ H(X) + \epsilon + \epsilon'} { \mu} + \frac{\max_i C_i}{q})(1 + \epsilon D).
			\end{equation}
			Choose $Q_3 = \ceilenv{\frac{8\max_i C_i}{\gamma}}$. When $q>\max\{Q_1,Q_2,Q_3\}$, we have
			\begin{equation}
			\label{total_cost_psi}
			\begin{aligned}
			&T(\Psi ) < (\frac{ H(\mathbf{X}) + \epsilon + \epsilon'} { \mu} + \frac{\max_i C_i}{q})(1 + \epsilon D)\\
			& =\frac{H(\mathbf{X})}{\mu} + (\frac{ \epsilon + \epsilon'} { \mu} + \frac{\max_i C_i}{q})(1 + \epsilon D) +\epsilon\frac{H(X)}{\mu} D\\
			&\leq \frac{H(\mathbf{X})}{\mu} + (\frac{\gamma}{8}+ \frac{\gamma}{8})(1+1)+\frac{\gamma}{2} = \frac{H(\mathbf{X})}{\mu} + \gamma.
			\end{aligned}
			\end{equation}
			As shown in~\cite[Theorem~3.2.1]{Cover}, for any $\gamma' > 0$, there exists a $Q_4>0$ such that when $q> Q_4$, the average codeword length per input symbol of $\Psi_1$ satisfies
			\begin{equation}
			\frac{1}{q}E[L(\Psi_1)] \leq H(\mathbf{X}) + \gamma'.
			\end{equation}
			The total  cost of $\Psi_2$ for binary uniform i.i.d. source is upper bounded by 
			\begin{equation}
			T'(\Psi_2) \leq \frac{\log_2 M}{\mu\ceilenv{q(H(\mathbf{X})+ \epsilon)+2}} + \frac{\max_{i}C_i}{\mu\ceilenv{q(H(\mathbf{X})+ \epsilon)+2}}.
			\end{equation}
			By an argument similar to that used to derive the upper bound on $T(\Psi)$ in (\ref{total_cost_psi}), we conclude that for any $\gamma'' > 0$, there exists a $Q_5 > 0$ such that when $q> Q_5$, 
			\begin{equation}
			T'(\Psi_2) \leq \frac{1}{\mu} + \gamma''.
			\end{equation}
			To summarize, given any $\gamma,\gamma',\gamma''>0$, for sufficiently large $q$, namely  $q>\max\{Q_1,Q_2,Q_3,Q_4,Q_5\}$, we can find a data compression encoder $\Psi_1$ such that 
			\begin{equation}
			\frac{1}{q}E[L(\Psi_1)] \leq H(\mathbf{X}) + \gamma'
			\end{equation}
			and a  code $\Psi_2$ for binary uniform i.i.d. source such that
			\begin{equation}
			T'(\Psi_2) \leq \frac{1}{\mu} + \gamma''.
			\end{equation}
			The concatenation of $\Psi_1$ and $\Psi_2$ will generate a code $\Psi = \Psi_2 \circ \Psi_1$ that has total   cost upper bounded by
			\begin{equation}
			T(\Psi) \leq \frac{H(\mathbf{X})}{\mu} + \gamma.
			\end{equation}
			This finishes the proof of the separation theorem.
			\begin{remark}
				Besides modified Varn coding, any fixed-to-variable (or fixed-to-fixed) coding scheme for a uniform source, such as the constant composition distribution matching codes introduced in~\cite{SchulteBocherer}, can be used to prove the separation theorem, as long as the cost of the codeword sequence is  bounded from above and the ratio of the upper bound to input sequence length is asymptotically optimal.
			\end{remark}
		\end{IEEEproof}
		
		\section{PROOF OF THEOREM~\ref{separation:type1}}
		\label{appen:type1}
		We know from  Theorem~\ref{thm:typeonetypetwo} that the problem of designing an optimal type-\Romannum{1} shaping code for channel with cost $\mathcal{C}$ and expansion $f_0$ is  related to designing an optimal type-\Romannum{2} shaping code for channel with cost $\mathcal{C}'$. The minimum total cost for channel with cost $\mathcal{C}'$ is $H(\mathbf{X})$ and the expansion factor for an optimal type-\Romannum{2} shaping code for channel with cost $\mathcal{C}'$ is $f_0' = f_0$. 
		
		We first fix some notation. Given a code $\phi$, denote by $T_{\mathcal{C}'}(\phi)$ the total cost of this code for channel with cost $\mathcal{C}'$. Denote by $f(\phi)$ its expansion factor and denote by $A_{\mathcal{C}}(\phi)$  and $A_{\mathcal{C}'}(\phi)$ the average cost of this code for channel with cost $\mathcal{C}$ and channel with cost $\mathcal{C}'$, respectively. The asymptotic symbol occurrence probability of $\phi$ is $\{P_i\}$ and $\hat{P}_i 
		= \frac{2^{-\mu C_i}}{N}$ is the symbol occurrence probability of the optimal type-\Romannum{1} shaping code.
		
		Before we prove the separation theorem for type-\Romannum{1} shaping code, we analyze the behavior of $f(\phi)$, $A_{\mathcal{C}}(\phi)$ and $A_{\mathcal{C}'}(\phi)$ when $T_{\mathcal{C}'}(\phi)$ approaches $H(X)$. The proof of Theorem~\ref{opt_expansion} determines the achievable total cost on channel with cost $\mathcal{C}'$. 
		\begin{figure}
			\centering
			\includegraphics[width=1\columnwidth]{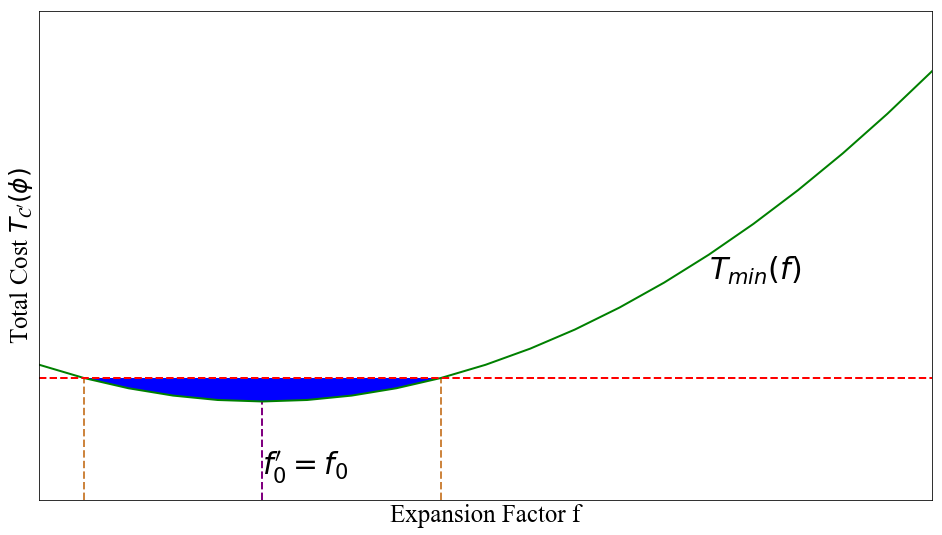}
			\caption{Achievable total cost vs. expansion factor}
			\label{fig:Achievablecostvsf}
		\end{figure}
		By combining equations~(\ref{equ:dfdmu}) and~(\ref{equ:dtdmu}), we have
		\begin{equation}
		\frac{\mathrm{d}T_\text{min}}{\mathrm{d}f} = -\frac{1}{\mu} \log_2 N.
		\end{equation}
		This implies that function $T_\text{min}(f)$ is continuous, strictly monotone decreasing on $[\frac{H(\mathbf{X})}{\log_2 |\mathcal{Y}|},f_0)$ and strictly monotone increasing on $(f_0,\infty)$.
		This is shown schematically by the  green curve $T_\text{min}(f)$ in Fig.~\ref{fig:Achievablecostvsf}.
		Thus for any $\zeta >0$, there exists a $\gamma$ such that if there exists a code such that $T_{\mathcal{C}'}(\phi)< H(\mathbf{X})+ \gamma$,  as indicated by the blue area in Fig.~\ref{fig:Achievablecostvsf}, then $|f(\phi)-f_0| < \zeta$. Such a code can always be found for $q$ sufficiently large, for example, by using the generalized Shannon-Fano construction (see~\cite{CsiszarKorner}).
		
		Now for any $\eta > 0$, $\zeta>0$, first choose $\zeta' = \min\{\frac{f_0}{2},\zeta, \frac{\eta\mu f_0^2}{4H(X)}\}$. Choose $\gamma'$ such that a type-\Romannum{2} shaping code with total cost upper bounded by $H(\mathbf{X})+ \gamma'$ has expansion factor $|f-f_0| < \zeta'$. Choose $\gamma = \min\{\gamma',\frac{\mu\eta f_0}{4}\}$, then $A_{\mathcal{C}'}(\phi)$  is bounded by
		\begin{equation}
		\begin{aligned}
		A_{\mathcal{C}'}(\phi) &= \sum_i P_i C_i' = \frac{T_{\mathcal{C}'}(\phi)}{f} < \frac{H(\mathbf{X})+\gamma}{f_0 - \zeta'} \\
		& = \frac{H(\mathbf{X})}{f_0} + \frac{H(\mathbf{X})}{f_0}\frac{\zeta'}{f_0-\zeta'} + \frac{\gamma}{f_0-\zeta'}\\
		&\leq \frac{H(\mathbf{X})}{f_0} + \frac{H(\mathbf{X})}{f_0}\frac{2\zeta'}{f_0} + \frac{2\gamma}{f_0}\\
		&\leq \frac{H(\mathbf{X})}{f_0} + \frac{\mu\eta}{2} +\frac{\mu\eta}{2} =  \frac{H(\mathbf{X})}{f_0} + \mu\eta.
		\end{aligned}
		\end{equation}
		Since the channel cost $C_i'$ is calculated by
		\begin{equation}
		C_i' = -\log_2 \hat{P}_i = -\log_2 \frac{2^{-\mu C_i}}{N} = \mu C_i + \log_2 N.
		\end{equation}
		$A_{\mathcal{C}}(\phi)$ is upper bounded by
		\begin{equation}
		\begin{aligned}
		A_{\mathcal{C}}(\phi) &= \sum_i P_i C_i = (\sum_i P_i C_i' - \log_2 N) / \mu\\
		&< (\frac{H(\mathbf{X})}{f_0 \mu} - \frac{\log_2 N}{\mu}) + \eta.
		\end{aligned}
		\end{equation}
		To summarize, for any $\eta >0$, $\zeta>0$, there exists a $\gamma >0$ such that if there exists a code $\phi$ such that $T_{\mathcal{C}'}(\phi) < H(\mathbf{X})+\gamma$, this code has expansion factor 
		\begin{equation}
		|f(\phi)-f_0| < \zeta
		\end{equation}
		and the average cost $A_\mathcal{C} (\phi)$ is upper bounded by
		\begin{equation}
		A_\mathcal{C} (\phi) < (\frac{H(\mathbf{X})}{f_0 \mu} - \frac{\log_2 N}{\mu}) + \eta.
		\end{equation}
		Now we present the proof of Theorem~\ref{separation:type1}.
		
		\begin{IEEEproof}
			We consider the channel with cost $\mathcal{C}$ and equivalent channel with cost $\mathcal{C}'$.
			Given $f,\eta,\zeta,\eta',\zeta',$ and $\gamma' >0$,  for $q$ sufficiently large, there exists a 
			data compression encoder $\Psi_1$ such that the average codeword length
			\begin{equation}
			\frac{1}{q}E[L(\Psi_1)] \leq H(\mathbf{X}) + \gamma'
			\end{equation}
			and a code $\Psi_2$ for a binary uniform i.i.d. source and costly channel with cost $\mathcal{C}'$ such that
			\begin{equation}
			T_{\mathcal{C}'}(\Psi_2) \leq 1 + \gamma''.
			\end{equation}
			The expansion factor of an optimal type-\Romannum{2} shaping code for binary uniform i.i.d. source and costly channel $\mathcal{C}'$ is
			\begin{equation}
			f' = \frac {f}{H(\mathbf{X})}.
			\end{equation}
			So  when $\gamma''$ is small enough, $\Psi_2$ has expansion factor 
			\begin{equation}
			|f(\Psi_2) - \frac{f}{H(\mathbf{X})}| < \zeta'
			\end{equation}
			and $A_\mathcal{C}(\Psi_2)$ is upper bounded by
			\begin{equation}
			\begin{aligned}
			A_\mathcal{C}(\Psi_2)& < (\frac{1}{\frac{f}{H(\mathbf{X})} \mu} - \frac{\log_2 N}{\mu}) + \eta' \\
			& = (\frac{H(\mathbf{X})}{f \mu} - \frac{\log_2 N}{\mu}) + \eta'.
			\end{aligned}
			\end{equation}
			The concatenation of $\Psi_1$ and $\Psi_2$ will generate a code $\Psi = \Psi_2 \circ \Psi_1$ that has total   cost upper bounded by
			\begin{equation}
			T_{\mathcal{C}'}(\Psi) \leq H(\mathbf{X}) + \gamma.
			\end{equation}
			When $\gamma$ is small enough, $\Psi$ has expansion factor 
			\begin{equation}
			|f(\Psi) - f| < \zeta
			\end{equation}
			and $A_{\mathcal{C}}(\Psi)$ is upper bounded by
			\begin{equation}
			A_{\mathcal{C}}(\Psi) < (\frac{H(\mathbf{X})}{f \mu} - \frac{\log_2 N}{\mu}) + \eta.
			\end{equation}
			This completes the proof.
		\end{IEEEproof}

	\end{appendices}
\end{document}